\documentclass[a4paper,11pt]{article}
\pdfoutput=1 

\usepackage{jcappub} 
\usepackage{verbatim}
\usepackage{comment}
\usepackage{float}
\usepackage{bm}
\usepackage{placeins}

\def\be{\begin{equation}}
\def\ee{\end{equation}}
\newcommand{\refeq}[1]{Eq.~(\ref{eq:#1})}          
  
\newcommand{\reffig}[1]{figure~\ref{fig:#1}} 
\newcommand{\reffigs}[2]{figures~(\ref{fig:#1})--(\ref{fig:#2})}
\newcommand{\refFig}[1]{Figure~\ref{fig:#1}}

\newcommand{\refsec}[1]{section~\ref{sec:#1}}          
          
\newcommand{\refapp}[1]{Appendix~\ref{app:#1}}

\newcommand{\refTab}[1]{Table~\ref{table:#1}}

\def\Mpch{\, h^{-1} \, {\rm Mpc}}
\def\iMpch{\,h\,{\rm Mpc}^{-1}}
\renewcommand{\d}{\delta}
\newcommand{\dbc}{\delta_{bc}}
\newcommand{\bdbc}{b_{\delta_{bc}}}
\newcommand{\Om}{\Omega_m}
\newcommand{\Oc}{\Omega_c}
\newcommand{\Ob}{\Omega_b}
\newcommand{\kmax}{k_{\rm max}}

\newcommand{\vx}{\mathbf{x}}
\newcommand{\vq}{\mathbf{q}}
\newcommand{\vk}{\mathbf{k}}

\title{Quantifying the impact of baryon-CDM perturbations on halo clustering and baryon fraction}
\author[a,b,c]{Hasti Khoraminezhad,}
\author[a,b,c]{Titouan Lazeyras,}
\author[d,e]{Raul E. Angulo,}
\author[f,g,h]{Oliver Hahn,}
\author[a,b,c,i]{Matteo Viel}
\affiliation[a]{SISSA, Via Bonomea 265, 34136 Trieste, Italy}
\affiliation[b]{INFN, Sezione di Trieste, Via Bonomea 265, 34136 Trieste, Italy}
\affiliation[c]{IFPU, Institute for Fundamental Physics of the Universe, via Beirut 2, 34151, Trieste, Italy}
\affiliation[d]{DIPC, Donosita International Physics Center, Paseo Manuel de Lardizabal, 4, 10018 Donosita-San Sebastian, Spain}
\affiliation[e]{IKERBASQUE, Basque Foundation for Science, Plaza Euskadi 5, 48013 Bilbao, Spain}
\affiliation[f]{Universit\'e Cote d'Azur, CNRS, Laboratoire Lagrange, Bvd. de l'Observatoire, CS 34229, 06304 Nice, France}
\affiliation[g]{Department of Astrophysics, University of Vienna, Türkenschanzstraße 17, 1180 Vienna, Austria}
\affiliation[h]{Department of Mathematics, University of Vienna, Oskar-Morgenstern-Platz 1, 1090 Vienna, Austria}
\affiliation[i]{INAF, Osservatorio Astronomico di Trieste, Via Tiepolo 11, 34143 Trieste, Italy}
\emailAdd{hkhorami@sissa.it, tlazeyra@sissa.it, reangulo@dipc.org,  oliver.hahn@univie.ac.at, viel@sissa.it}

\abstract{Baryons and cold dark matter (CDM) did not comove prior to recombination. This leads to differences in the local baryon and CDM densities, the so-called baryon-CDM isocurvature perturbations $\delta_{bc}$. These perturbations are usually neglected in the analysis of Large-Scale Structure data but taking them into account might become important in the era of high precision cosmology. Using gravity-only 2-fluid simulations we assess the impact of such perturbations on the dark matter halos distribution. In particular, we focus on the baryon fraction in halos as a function of mass and large-scale $\dbc$, which also allows us to study details of the nontrivial numerical setup required for such simulations. We further measure the cross-power spectrum between the halo field and $\delta_{bc}$ over a wide range of mass. This cross-correlation is nonzero and negative which shows that halo formation is impacted by $\delta_{bc}$. We measure the associated bias parameter $b_{\d_{bc}}$ and compare it to recent results, finding good agreement. Finally we quantify the impact of such perturbations on the halo-halo power spectrum and show that this effect can be degenerate with the one of massive neutrinos for surveys like DESI.}


\begin{document}
\maketitle
\flushbottom


\section{Introduction}
\label{sec:intro}

The current standard paradigm in cosmology is that structure formation finds its source in quantum fluctuations generated and amplified during Inflation. These fluctuations then grew under the action of gravity down to low redshift forming the observed galaxies and galaxy clusters. Depending on the details of the inflationary scenario (including reheating), baryons and cold dark matter (CDM) can exhibit different fluctuations at the end of Inflation, giving rise to isocurvature density perturbations (i.e. relative perturbations in the baryon and CDM density without affecting the total matter potential), see \cite{Polarski:1994rz,Linde:1996gt,Liddle:1999pr,Langlois:2000ar,Notari:2002yc,Lyth:2002my,Ferrer:2004nv,Li:2008jn,Grin:2011,Valiviita:2012ub,Huston:2013kgl,Christopherson:2014eoa,He:2015msa} and references therein\footnote{Notice that single field Inflation does not generate such perturbations}. Furthermore, regardless of the details of Inflation, such perturbations are generated in the early Universe due to the coupling between baryons and photons through Compton scattering prior to recombination. This coupling does not apply to CDM hence making the spatial and velocity distributions of the baryons and CDM very different just after decoupling. 

Gravitational evolution after recombination slowly erases this difference since baryons can then fall in CDM potential wells. This process is normally assumed to be finished before redshift zero implying that the distributions of the two fluids become identical on large scales, with the power spectrum of the fluctuations of each fluid given by the total matter power spectrum. This assumption is however not exactly true, and recently substantial effort was put in correctly describing and simulating the evolution of the two fluids across cosmic history. In this process, so-called \textit{2-fluid simulations} where the baryons and CDM fluid are initialized with different transfer functions have become an essential tool \cite{Yoshida:2003,OLeary:2012,Angulo:2013qp,Bird:2020,Hahn:2020,Michaux:2020,Rampf:2020}. In this paper we focus on the \textit{gravity-only} version of such simulations, i.e. we neglect the late-time impact of baryonic processes by introducing hydrodynamic physics in the simulations, which is justified as we will primarily study large scales.

The difference in the baryon and CDM perturbations is also expected to affect structure formation \cite{Ahn:2016bcr} and the clustering of Large-Scale Structure (LSS) tracers \cite{Schmidt:2016,Chen:2019cfu}. It might then be important to incorporate them in the analysis of LSS data in order to obtain unbiased cosmological constraints (see \cite{Heinrich:2019sxl} for example). In the scope of effective-field theory, the overdensity of tracers $\d_h$, such as dark matter halos, is expressed in terms of operators $O$ constructed out of the \textit{total} matter density field $\d_m$ and tidal field $K_{ij}$, multiplied by numerical coefficients $b_O$, the bias parameters (see \cite{Desjacques:2016bnm} for a very complete review)
\be
\d_h(\vx, \tau) = \sum_O b_O(\tau) O(\vx,\tau).
\label{eq:generalbiasexp}
\ee
To take into account the impact of relative perturbations between the baryon and CDM fluids, one must add new terms proportional to the difference between the baryon density $\d_b$ and the CDM one $\d_c$. As we show in \refsec{theory} two new terms appear at linear order already: a constant mode $\dbc$, and a decaying one proportional to the difference between the velocity divergence of the two fluids $\theta_{bc}$. These terms are multiplied by their respective bias parameters $\bdbc$ and $b_{\theta_{bc}}$. Previous works have attempted to estimate and measure these terms \cite{Barkana:2011,Schmidt:2016,Barreira:2019,Chen:2019cfu,Hotinli:2019wdp}. In particular, \cite{Barreira:2019} used an extension of the so-called separate universe simulations (e.g. \cite{McDonald:2001fe,Wagner:2014aka} and references therein) to give the first precise measurement of $\bdbc$.

Large-scale isocurvature perturbations may also affect the baryon content of halos. This is important since the fraction of baryons in halos, $F_b$, can be used to e.g. determine cosmological parameters such as $\Omega_b h^2$ or $\Omega_m$ (e.g. \cite{Lubin:1995er,Sadat:2001en}), constraint primordial non-Gaussianity \cite{Maio:2011av} or infer cluster and halo mass from the Tully-Fisher relation \cite{McGaugh:1997ej,Ilic:2015wby}. Furthermore, $F_b$ directly determines how much material is available for galaxy formation, hence affecting one of the main luminous tracers of LSS. The baryon fraction in halos and clusters was measured in various manners in e.g. \cite{Papastergis:2012wh,Baghram:2017qby}, as well as its variations on large-scales \cite{Holder:2009gd}. On the numerical side, it was also extensively studied in e.g. \cite{He:2005hb,Ettori:2005hp,Kravtsov:2005ab,Crain:2006sb,Gottloeber:2007eb} in the scope of hydrodynamic simulations (i.e. focusing on late-time baryonic physics and radiative effects), investigating its departure from the cosmic mean. Here we focus again on gravity-only 2-fluid simulations only in order to assess the impact of early-time baryonic physics on this quantity, which was never done before to our knowledge. This also allows us to test the limits of our numerical setup (which is nontrivial for such simulations) and our ability to simulate correctly the distribution of particles on small scales.

The goal of this paper is to investigate how baryon-CDM isocurvature perturbations impact halo clustering using gravity-only 2-fluid simulations. In particular we focus on the baryon fractions in halos $F_b$, and the baryon-CDM perturbation bias $\bdbc$. We also measure the baryon-CDM cross-power spectra and halo-bc one, showing it to be nonzero which confirms the impact of baryon-CDM perturbations on halo clustering. Importantly, the technique used in this paper is completely different from that of \cite{Barreira:2019}, and our simulations encompass additional physical processes to generate $\bdbc$ beyond inflationary ones, since we modify the baryon and CDM power spectra separately, keeping the total matter power spectrum fixed on all scales. This allows for an independent measurement of this little studied parameter and its importance in the bias expansion.

This paper is organized as follow. In \refsec{theory} we give a brief overview of isocurvature perturbations (\refsec{bcdmpert}) and how to measure $\bdbc$ (\refsec{measurebdbc}). We then turn to a detailed description of our simulations in \refsec{sims}. We start by describing how we generate initial conditions in \refsec{ics} and we give a few details of the simulations in \refsec{nbody}. We then turn to various numerical tests to validate our setup (\refsec{numtests}), and we describe the halo finding procedure in \refsec{halofinding}. We present our results in \refsec{results}, focusing first on the baryon fraction in \refsec{fb}, and halo bias and power spectra in \refsec{2fPSbias}. We conclude in \refsec{conclusion}. The appendices present a rapid overview of the separate universe technique used in \cite{Barreira:2019} to measure $\bdbc$ in \refapp{baryonCDMSU}, and additional numerical tests of the 2-fluid simulations in \refapp{tests}.


\section{Theory}
\label{sec:theory}

\subsection{Baryon-CDM pertubations}
\label{sec:bcdmpert}

In this section we summarize how baryon-CDM perturbations are generated in the early Universe, and we review the formalism to derive their evolution. All this was already discussed in details in \cite{Schmidt:2016,Barreira:2019} so we stay concise and refer the interested reader to these papers.

We restrict ourselves to linear perturbation theory since we will focus on (very) large scales in this work. 

We start by writing the Euler and Continuity equations for the CDM and baryon components after decoupling. We express these in terms of the total and relative density perturbations
\be
\d_m=f_b \d_b + (1-f_b)\d_c, \quad \d_r=\d_b-\d_c\, ,
\label{eq:deltas}
\ee
where $\d_b$ and $\d_c$ are the baryon and CDM fractional density perturbations and $f_b=\Ob/\Om$, which yields
\begin{align}
\frac{\partial^2}{\partial \tau^2}\d_m + \mathcal{H}\frac{\partial}{\partial \tau}\d_m-\frac32\Om(a)\mathcal{H}^2\d_m & =0, \nonumber \\
\frac{\partial^2}{\partial \tau^2}\d_r + \mathcal{H}\frac{\partial}{\partial \tau}\d_r & =0.
\label{eq:cont-eul}
\end{align}
As shown in \cite{Schmidt:2016} these two equations in term of these variables admit the following solutions 
\begin{align}
\d_m(\tau) & =A_+ D_+(\tau) + A_- H(\tau), \nonumber \\
\d_r(\tau) & =R_+  + R_- D_r(\tau),
\label{eq:sols}
\end{align}
where $A_\pm$, $R_\pm$ are constants, $D_+(\tau)$ is the usual linear matter growth rate, and $D_r(\tau)$ can be approximated to $-2a^{-1/2}(\tau)$ during matter domination. Apart from the usual growing and decaying modes of $\d_m$, we are interested in the two modes of $\d_r$. The first one is a constant mode $\dbc$\footnote{We use the subscript $m$ for total matter and reserve $bc$ for the difference $b-c$.} of compensated perturbations with $\d_m=0$, i.e. $\delta\rho_b=-\d\rho_c \Leftrightarrow f_b\d_b=-(1-f_b)\d_c$, but $\d_r \neq 0$. The second mode is a decaying one that can be shown to be related to the divergence of peculiar velocity perturbations between the two fluids, $\theta_{bc}=\theta_b-\theta_c$ (see \cite{Schmidt:2016}). We hence get
\be
\d_r(\vx,\tau)=\dbc(\vx)+\frac{\theta_{bc}(\vx,z=0)}{H_0}D_r(\tau).
\label{eq:dr}
\ee

These two new perturbation terms must enter the bias expansion at linear order already with new associated bias parameters. They should normally be evaluated at the Lagrangian position $\vq(\vx)$ corresponding to Eulerian position $\vx$. However at linear order we can neglect this which allows us write the fractional halo density perturbation at linear order as
\be
\d_h(\vx,\tau)=b_1(\tau)\d_m(\vx,\tau) + b_{\dbc}(\tau)\dbc(\vx)+b_{\theta_{bc}}(\tau)\theta_{bc}(\vx,\tau).
\label{eq:biasexp}
\ee


\subsection{Measuring $b_{\d_{bc}}$}
\label{sec:measurebdbc}

We now turn to expressions for the halo cross-power spectra $P_{hm}$ and $P_{hbc}$. As argued in \cite{Schmidt:2016,Barreira:2019} the two last terms in \refeq{biasexp} should be much smaller than the first one, which is why they are normally neglected. Furthermore, the last term proportional to $\theta_{bc}$ is expected to be much smaller than the second one proportional to $\dbc$. We will hence neglect the velocity divergence term in what follows.

The cross-power spectrum $P_{hbc}$ is defined as 
\be
(2 \pi)^3 \delta_D(\vk+\vk') P_{hbc}(k) = \left\langle \d_h(\vk) \dbc(\vk') \right\rangle,
\label{eq:phbc}
\ee
and similarly for $P_{hm}$. Plugging \refeq{biasexp} into \refeq{phbc} we get
\begin{align}
P_{hm}(k) & = b_1 P_{mm}(k) + b_{\dbc} P_{mbc}(k), \nonumber \\
P_{hbc}(k) & = b_1 P_{mbc}(k) + b_{\dbc} P_{bcbc}(k),
\label{eq:powerspectra}
\end{align}
where $P_{mm}$ is the usual matter power spectrum, while $P_{mbc}$ and $P_{bcbc}$ are the cross- and auto-power spectra of $\dbc$ with the matter field and itself respectively. Notice that if we neglect the last term in the second line, the ratio $P_{hbc}/P_{mbc}$ should go to $b_1$. From these two equations we can get expressions for $b_1$ and $b_{\dbc}$ in the low $k$ limit
\begin{align}
b_1 & = \lim_{k\rightarrow 0} \frac{P_{hm}(k) - b_{\dbc} P_{mbc}(k)}{P_{mm}(k)}, \nonumber \\[10pt] 
b_{\dbc} & = \lim_{k\rightarrow 0} \frac{P_{hbc}(k)P_{mm}(k)- P_{hm}(k)P_{mbc}(k)}{P_{bcbc}(k)P_{mm}(k)-P_{mbc}^2(k)}.
\label{eq:bias}
\end{align}
We can hence get $b_{\dbc}$ from the second of these two equations and insert it in the first one to get $b_1$. It is very clear that the term proportional to $b_{\dbc}$ in the first line of \refeq{bias} represents the deviation of $b_1$ from the traditional ratio $P_{hm}/P_{mm}$ due to baryonic effects in 2-fluids simulations.

Actually, a more practical way to obtain $b_{\dbc}$ with more constraining power is to obtain $b_1$ by the usual ratio $P_{hm}/P_{mm}$ at low $k$ in 1-fluid simulations, and to subtract $b_1 P_{mm}$ from the first line of \refeq{powerspectra} in order to detect any deviation in the usual relation in 2-fluids simulations, i.e
\be
b_{\dbc} = \lim_{k\rightarrow 0} \frac{P_{hm}(k)-b_1^{1 {\rm f}}P_{mm}(k)}{P_{mbc}(k)},
\label{eq:bdbc1f}
\ee
where we have used the superscript ``1f'' to denote the linear bias as measured in 1-fluid simulations, i.e.
\be
b_1^{1 {\rm f}}=\lim_{k\rightarrow 0}\frac{P_{hm}^{1 {\rm f}}(k)}{P_{mm}^{1 {\rm f}}(k)}.
\label{eq:b11f}
\ee
Finally we could do the same reasoning adding $\theta_{bc}$. However we expect this term to be subdominant since it is a decaying one, and adding it would probably only diminish our constraining power on $\bdbc$. We will however compare our results to ones obtained  with the 1-fluid ``separate universe simulations'' technique of \cite{Barreira:2019} as outlined in \refapp{baryonCDMSU}. Since our 2-fluid simulations include by default $\theta_{bc}$, which is not the case of the 1-fluid ones, any difference between the two measurements can be attributed to our neglection of $\theta_{bc}$ and would hence be an estimation of the magnitude of this term.


\section{Simulations and halo finding}
\label{sec:sims}

We now turn to a detailed description of our set of simulations. As was discussed in \cite{Angulo:2013qp,Bird:2020,Hahn:2020}, obtaining the correct evolution of each species in 2-fluid simulations, even at linear order only, is already a nontrivial task which is why we thoroughly present a number of tests to validate our setup in this section. For the sake of shortness we present less important tests results in \refapp{tests}.

Our fiducial cosmology is consistent with the Planck 2018 one \cite{Aghanim:2018eyx}, detailed as follows: $\Om=0.3111$, $\Omega_{b}=0.0490$, $\Omega_{c}=0.2621$, $\Omega_{\Lambda}=0.6889$, $n_{s}=0.9665$, $\sigma_{8}=0.8261$ and  $h=0.6766$. The box size is $L_{box}=250 \, \Mpch$ on each side for all our simulations. 

We run two sets of simulations. The first one is a standard gravity-only one with one species of particles that we refer to as ``1-fluid''. In addition to the fiducial cosmology we run two additional cosmologies with enhanced (``High'') and lowered (``Low'') $\Omega_b$ (while adapting $\Oc$ to keep $\Om$ fixed) in order to compute $\bdbc$ in the same fashion as \cite{Barreira:2019}. We also use the fiducial simulation of this set to compute the linear bias $b_1$. Explicitly speaking, these simulations have only CDM particles, and we use $512^3$ mass elements. The way we run this set of ``separate universe simulations of baryon-CDM perturbations'' is identical to what was done in \cite{Barreira:2019} and we refer the reader to their paper for more details.

The second set of simulations contains two distinct fluids representing baryons and CDM each with different primordial density and velocity fluctuations. Explicitly, we use two different transfer functions to initialize the two fluids that we then evolve jointly. Each fluid consist of $512^3$ mass elements. We refer to this second set as ``2-fluid''. Furthermore we ran a hybrid version of the 2-fluid simulation where the two fluids are initialized with the same transfer functions, in order to check our numerical setup. We insist that we do not include any hydrodynamical effects for baryon evolution, and that all our simulations are gravity-only. \refTab{simu} summarizes the varying parameters of our simulations. In the following sections we go into the details of our numerical setup for 2-fluid simulations and present some sanity checks.

\begin{table}
\centering
\begin{tabular}{ |c|c|c|c|c|c|c|c|c| } 
 \hline
 Name  & $N_b$ & $N_c$ & $m_b$ & $m_c$ & $\Omega_c$ & $\Omega_b$ & $N_{\rm real}$ & TFs \\ 
 \hline \hline
 1-fluid Fid & 0 & $512^3$ & -- & $1.0051$ & 0.2621 & 0.049 & 16 & -- \\ 
 \hline
 1-fluid High & 0 & $512^3$ & -- & $1.0051$ & 0.2596 & 0.0515 & 16 & -- \\ 
 \hline
 1-fluid Low & 0 & $512^3$ & -- & $1.0051$ & 0.2645 & 0.0466 & 16 & -- \\ 
 \hline
 2-fluid-diff & $512^3$ & $512^3$ & $0.1583$ & $0.8468$ & 0.2621 & 0.049 & 4 & 2 \\ 
 \hline
 2-fluid-same & $512^3$ & $512^3$ & $0.1583$ & $0.8468$ & 0.2621 & 0.049 & 4 & 1 \\ 
 \hline
\end{tabular}
\caption{Summary of our sets of simulations. All simulations are gravity-only and have a box size of $250 \Mpch$ on each side. $N_b$ and $N_c$ are the number of baryonic and CDM particles respectively. $m_b$ and $m_c$ are the corresponding mass in units of $10^{10} M_\odot/h$. $N_{\rm real}$ corresponds to the number of realizations we ran of each simulations in order to build statistics, and ``TFs'' refers to the number of transfer functions used to initialize the two fluids (1 means that they are initialized with the same transfer function corresponding to the weighted total matter one).}
\label{table:simu}
\end{table}

\subsection{Initial Conditions}
\label{sec:ics}

We generate the initial conditions (initial position and velocities) for the particles in our simulations at an initial redshift $z_{i}=49$ using the publicly available MUSIC code \cite{Hahn_2011}. 

The 1-fluid simulations are initialized in the standard way by computing the primordial matter power spectrum using the CAMB code \cite{Lewis_2000} at $z=0$, and back-scaling it to the initial redshift assuming growing mode only for the specified cosmology. On the other hand, in the case of 2-fluid simulations we compute the transfer functions for baryons and CDM from CAMB directly at $z=49$. Notice that the total matter power spectrum is the same for both 1-fluid and 2-fluid simulations.  

We then compute the displacement and velocity fields using the Zel'dovich approximation \cite{Zeldovich-1970} (for simplicity we are not using the second-order Lagrangian Perturbation Theory (LPT) formalism here; notice however that it was recently figured out for two fluids in \cite{Rampf:2020}). Furthermore we used the fixed mode amplitude technique incorporated in MUSIC of \cite{Angulo:2016hjd}, in which the modulus of the white noise Fourier modes is set to unity in order to suppress the impact of cosmic variance. 


\subsection{N-body simulations details}
\label{sec:nbody}

Our simulations are performed with the cosmological code Gadget-2 \cite{Springel:2005mi} with a numerical setup very similar to the one of \cite{Angulo:2013qp}. As stated before, we compute only gravitational interactions and neglect all hydrodynamical effects, implying that baryons behave like a collisionless fluid. This is because we are only interested in investigating the effect of baryon isocurvature perturbations generated in the early Universe.

As discussed in detail in \cite{Angulo:2013qp}, one of the issues that needs to be addressed in these kind of collisionless simulations is the force resolution for the light fluid. Indeed a too high force resolution for the mass resolution could cause a spurious coupling between CDM and baryons affecting their clustering features and the growth of structures on all scales. A simple solution to remedy that is to make the baryon smoothing length unusually high. In fact, as shown in \cite{Angulo:2013qp}, the force softening must be of the order of the mean baryon inter-particle distance in order to recover the correct linear evolution. However, this is a problem since this length can become of the order of $1 \Mpch$ and small halos can have a final radius smaller than this implying that we would not simulate correctly structure formation at the small mass end of the mass function.

Another solution discussed in \cite{Angulo:2013qp}, that we use in this work, is to use the  adaptive gravitational softening (AGS) \textit{for baryons only} \cite{Iannuzzi2011AdaptiveGS}, implemented in Gadget. This technique allows the softening length to vary in space and time according to the density of the environment. Specifically, in our case the force acting on baryonic particles is softened adaptively using an SPH kernel with a size set by the 28 closest neighbours (DesNumNgb=28 in Gadget). Furthermore we set a floor for the minimum softening length $\epsilon=12.5 h^{-1} {\rm kpc}$ corresponding to roughly $1/40$-th of the mean inter-particle separation of baryons. The CDM softening length is kept constant through space and time to $\epsilon=12.5 h^{-1} {\rm kpc}$, also corresponding to $1/40$-th of the mean CDM inter-particle separation. We present validating tests of this setup in the next section as well as the effect of varying specific details of the force softening in \refapp{tests}.

Finally let us note that recently several papers tackled the issue of the spurious coupling between the light and heavy particles without introducing a large softening length. Ref. \cite{Bird:2020} claimed that this can be done by using a Lagrangian glass for the baryon particles. The recent papers \cite{Hahn:2020} and \cite{Rampf:2020} formally generalized LPT to an arbitrary order $n$ and use variations in particle masses to resolve the spurious deviations from expected perturbative results in baryon-CDM simulations. While we do not attempt to compare rigorously our setup with theirs, we compare the results for the baryon fraction $F_b$ when using our one or the one of \cite{Hahn:2020} in \refsec{fb}.


\subsection{Numerical tests}
\label{sec:numtests}

In this section, we present numerical tests to validate our 2-fluid simulations. For the sake of shortness we restrict ourselves to a few key checks and present additional numerical tests in \refapp{tests}. All the tests presented here concern the 2-fluid-diff-TF set of simulations.

\begin{figure}
\centering
\includegraphics[scale=0.45]{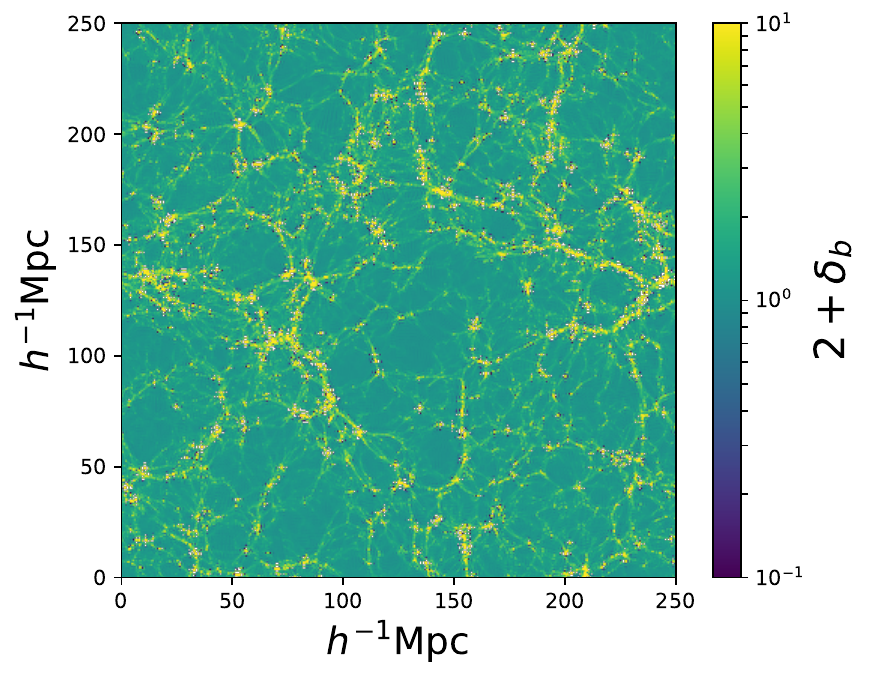}
\includegraphics[scale=0.45]{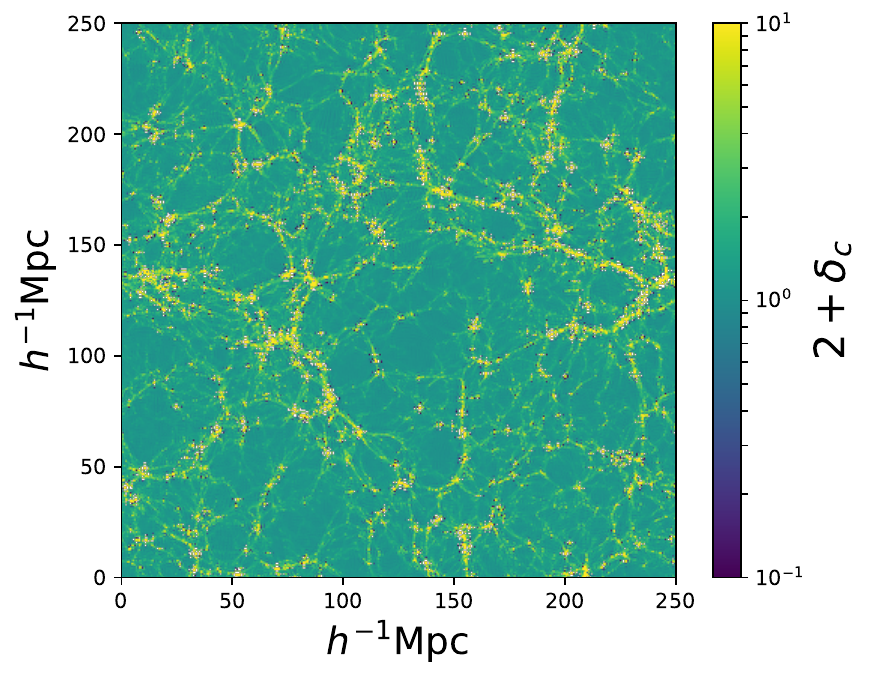}
\includegraphics[scale=0.45]{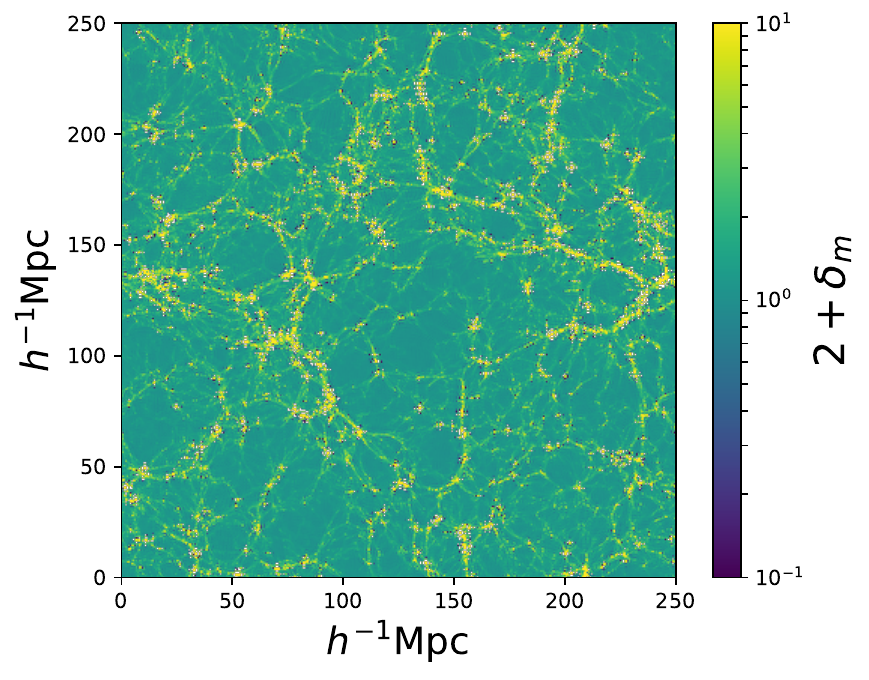}
\includegraphics[scale=0.45]{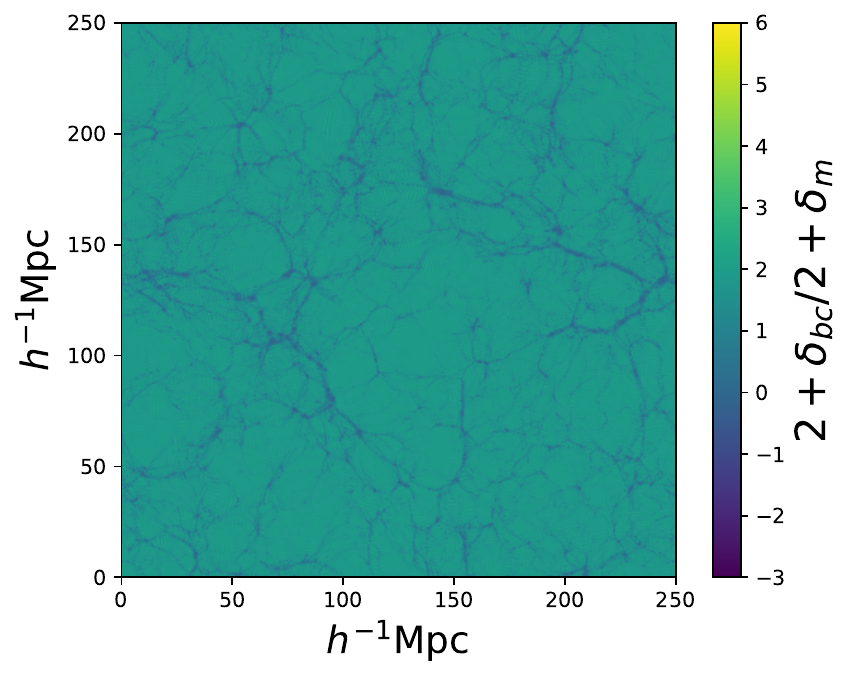}
\caption{Maps of density fields in a slice of thickness $10$ $\Mpch$. The top left panel shows the baryon density field while the top right panel shows the CDM one. In the bottom left panel we present the total matter density defined as in \refeq{deltas}. The bottom right one shows the relative baryon-CDM field denoted as $\delta_{bc}=\delta_b-\delta_c$ (if $\theta_{bc} \rightarrow 0$), normalized to $\d_m$. We show this ratio to allow for better visualization. We see that baryons do indeed trace the CDM fluid but with a small lag which makes $\dbc$ negative in high $\d_m$ regions, while it is positive in low density ones.}
\label{fig:density}
\end{figure}

We start with the maps of the density fields in a slice of thickness $10\, h^{-1} \rm Mpc$ at $z=0$ presented on \reffig{density}. The top left panel represents the baryon density filed while the top right panel shows the CDM density field. Here one can see how baryons are following the CDM. In bottom the left panel we present the total matter density defined as $\delta_{m}=f_b\delta_{b}+(1-f_{b})\delta_c$. The bottom right panel shows the relative baryon-CDM field denoted as $\delta_{bc}=\delta_b-\delta_c$ (if $\theta_{bc} \rightarrow 0$), normalized to $\d_m$ (to allow for better visualization). This figure gives us a visual validation of our simulations by confirming the presence of the usual structures in the cosmic web, the fact that baryons closely trace CDM, and allows us to visualize the relative baryon-CDM field showing that it is smaller in amplitude than the total one and has a negative value in average. Furthermore we see that baryons trace CDM with a small lag which makes $\dbc$ negative in high $\d_m$ regions, while it is positive in low density ones. 

\begin{figure}
\centering
\includegraphics[scale=0.4]{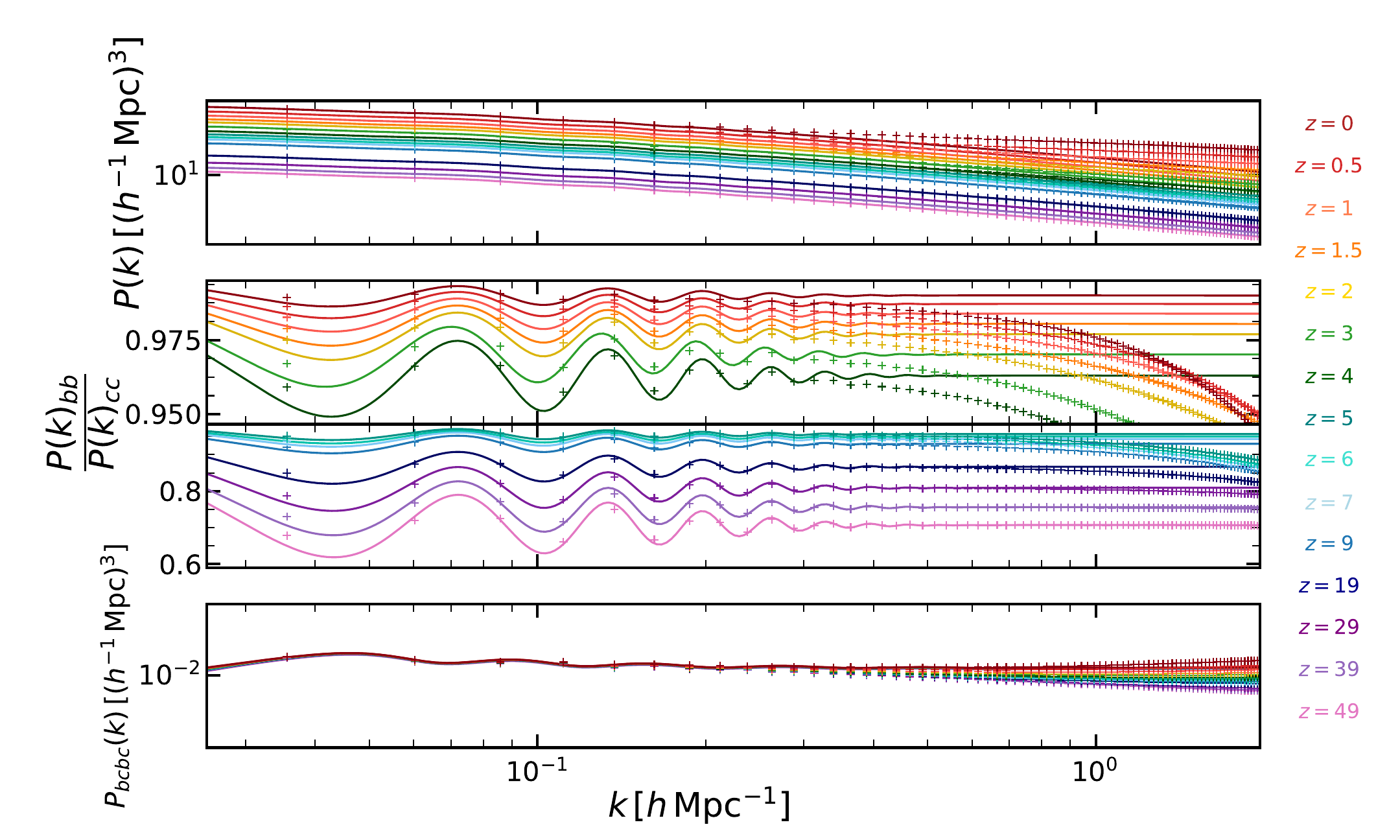}
\caption{Power spectra as a function of wavenumber $k$ for 14 different redshifts indicated by the color coding. Solid lines show the prediction of the linear perturbation theory and plus (``+'') markers represent the results from simulations. The top panel presents the growth of the total matter field, the two middle ones show the ratio between the measured baryon to CDM power spectra, and the bottom one presents results for the relative perturbation $\d_{bc}$ auto-power spectrum. We see that measurements agree with linear theory up to $k\sim 0.3 \iMpch$ down to $z=0$, as is expected. On the middle panels we see the BAO wiggles due to the fact that they are present only in the baryon power spectrum but not in the CDM one. We see that the difference between the baryon and CDM power spectra becomes of the order of $1\%$ at $z=0$. This is the impact of this difference on dark matter halos that we want to study in this work. The clear suppression of baryon perturbations compared to the CDM ones on small scales is due to our treatment of the force softening with AGS for baryons. We see no redshift dependency of $P_{bcbc}$ on large scales on the fourth panel as is expected since the relative density $\d_r$ can quickly be approximated by the constant mode $\d_{bc}$ as $\theta_{bc}\rightarrow 0$ as explained in section \refsec{bcdmpert}. The good agreement between our measurements and linear theory for all curves validates our numerical setup.}
\label{fig:PbboverPc}
\end{figure}

We now turn to comparing the measured power spectra of the density fields shown on \reffig{density} with linear theory at different redshifts. The measurements were performed by mapping the particle distribution using a cloud-in-cell (CIC) scheme on to a $1024^{3}$ grid and then Fast Fourier Transforming the field. The top panel of \reffig{PbboverPc} compares the measured growth of the total matter power spectrum (``+'' signs) with the prediction of the linear perturbation theory, shown as solid lines, for different redshift. We can see that our simulations reproduces the expected linear growth from $z_{i}=49$ up to $z=0$, up to $k \sim 0.3 \iMpch$ at $z=0$. On small scales and at low redshift, the nonlinear growth of structures dominates which is why the measured power spectrum becomes higher than the linear one. This agreement supports the correctness of our numerical calculations. The two middle panels show the ratio between the baryon and CDM power spectra at various redshifts. Again measurements are represented by ``+'' while the solid lines show the linear prediction. The fact that the solid curves are systematically different from unity implies that the overall shape of the power spectrum of baryons and CDM is different even on relatively large scales. The density perturbations in the baryon density field is smaller than the dark matter density field at all redshifts, this is because of the extra suppression produced by radiation pressure before recombination. In addition we clearly see the BAO wiggles due to the fact that BAOs are only present in the baryon power spectrum. These wiggles become less and less important at lower redshift as gravitational evolution slowly washes them away. We see that the difference between the two power spectra becomes of order $1\%$ at redshift zero which is why simulations are usually initialized assuming the same power spectra for the two fluids. It is however precisely the impact of this difference on dark matter halos that we want to investigate in this paper. The clearly visible suppression in $P_{bb}$ compared to $P_{cc}$ on small scales is due to our treatment of the force softening with AGS for baryons. Finally, the fourth panel of \reffig{PbboverPc} presents the evolution of the relative baryon-CDM density field auto-power spectrum $P_{bcbc}$ obtained from evaluating $\left\langle \delta_{bc}(\vk),\delta_{bc}(\vk^{\prime}) \right\rangle$ with $\delta_{bc} \approx \delta_{b} - \delta_{c}$ (simulation results are once again shown by ``+'' markers and the solid lines comes from the linear perturbation theory by  the CAMB code). We can see that there is no redshift dependency of $P_{bc\,bc}$ which is expected since, as we showed in \refsec{bcdmpert}, the term proportional to $\theta_{bc}$ rapidly decays in $\delta_r$ leaving only the term $\d_{bc}$ which is constant in time. The departure from this behaviour on small scales is probably due to nonlinear evolution as well as errors in our numerical setup due to the AGS. The results presented on this figure represent nontrivial tests of our simulations, and the good agreement with linear theory up to $k\sim0.3 \iMpch$ at $z=0$ allows us to validate them.


\subsection{Halo finding}
\label{sec:halofinding}

Halos were identified using the Amiga Halo Finder (AHF) \cite{Gill:2004km, Knollmann2009AhfAH}, which identifies halos as spherical overdensities (SO) in the spatial distribution of particles in the simulations. The virial radius is defined as the radius within which the average density is given by $\bar{\rho}_{vir}(z)= \Delta_{m}(z) \, \rho_{m}(z)$ where $\rho_m$ is the total matter background density (i.e. we identified halos using both baryons and CDM particles), and we choose the overdensity threshold $\Delta_m=200$. We refer the interested reader to \cite{Knollmann2009AhfAH} for more details. We set the minimum number of particles per halo to $20$ and we use only main halos in this work. Finally, we identified halos at $z=0, \, z=0.5, \, z=1, \, z=1.5, \, z=2,$ and $z=3$. We bin the mass range of halos in $8$ tophat bins of width $0.5$ in logarithmic scale centered from $\log M = 11.20$ to $\log M = 14.70$, where $\log$ is the base $10$ logarithm, to ensure that we have enough halos in each bin.

\begin{figure}
\centering
\includegraphics[scale=0.35]{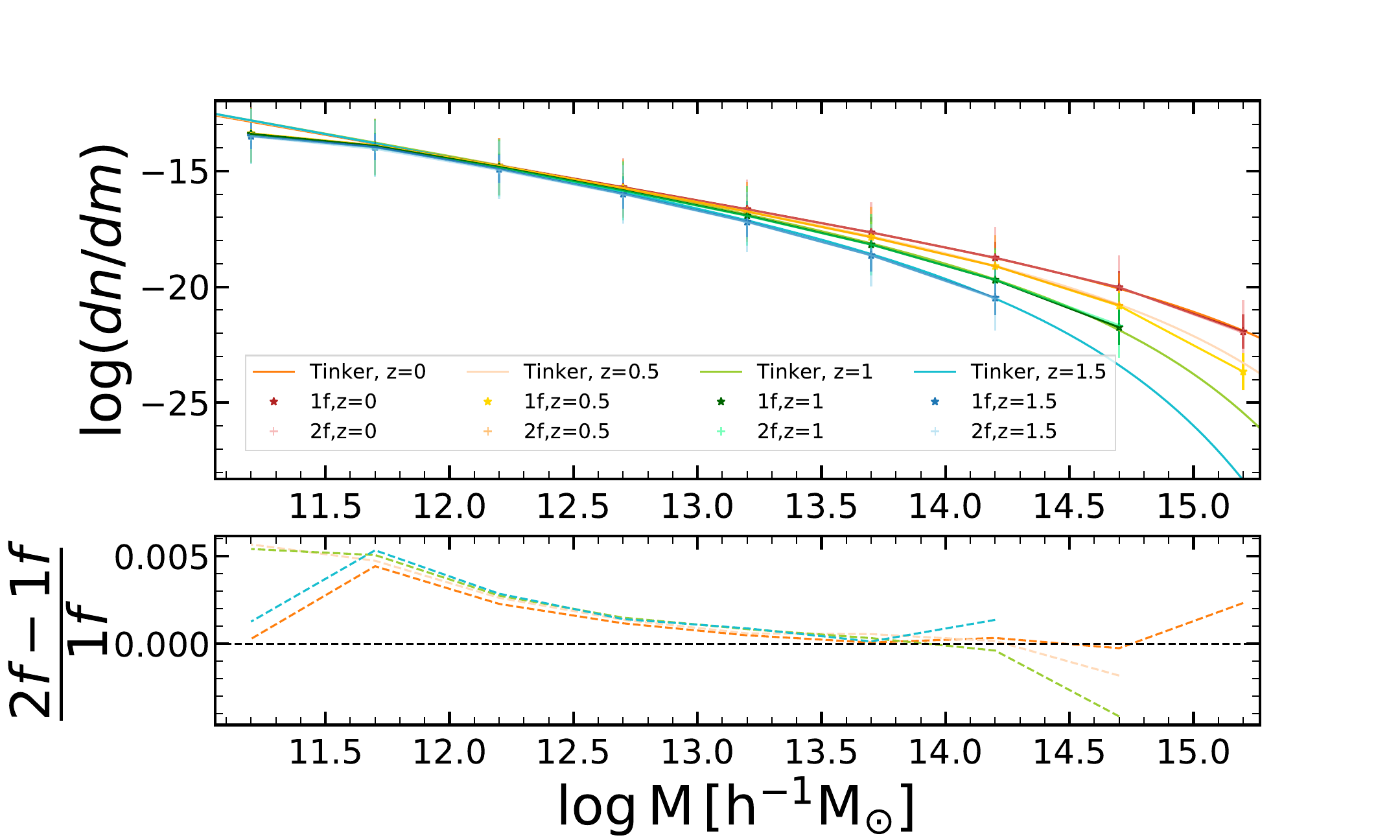}
\caption{Halo mass functions for our 1-fluid and 2-fluid simulations at $z=0, \,$ $z=0.5$, $z=1$ and $z=1.5$ indicated by the color coding. The symbols present our measurements while the solid lines are the Tinker 08 mass function. The shaded region show the $1\sigma$ errorbars obtained as the error on the mean over all realizations for each simulation. The lower panel shows the relative difference between the 1 and 2-fluid cases. We find good agreement with the Tinker fit as well as a subpercent difference between the two simulations sets as expected.}
\label{fig:dndM}
\end{figure}

To validate our simulations and our halo finding we present results for the halo mass function, ${\rm d}n(z)/{\rm d}M$ - i.e. the number of halos per unit volume per unit mass at redshift $z$ - for our 1-fluid and 2-fluid simulations and a comparison with the well known Tinker halo mass function \cite{Tinker:2008ff} on \reffig{dndM}. We present results at $z=0, \,$ $z=0.5$, $z=1$ and $z=1.5$ indicated by the color coding. The symbols present our measurements while the solid lines are the Tinker 08 mass function. The shaded region show the $1\sigma$ errorbars obtained as the error on the mean the over all realizations for each simulation. The lower panel shows the relative difference between the 1 and 2-fluid cases. We find good agreement with the Tinker fit, validating our halo catalogs, as well as a subpercent difference between the two simulations sets as expected since early baryonic effects should be at maximum of order $1\%$ at low redshift. Furthermore since we keep the total matter power spectrum fixed on all scales we do not expect the halo abundance to depend strongly on the isocurvature perturbations.


\section{Results}
\label{sec:results}

In this section we present results for the baryon fraction in halos, as well as the 2-fluid cross- and auto-power spectra constructed from $\delta_{m}$, $\delta_{bc}$ and $\delta_{h}$. We also show measurements of $b_{\delta_{bc}}$ as a function of halo mass $M$ and linear bias $b_1$. 

\subsection{Mean baryon fraction in halos}
\label{sec:fb}

\begin{figure}
\centering
\includegraphics[scale=0.3]{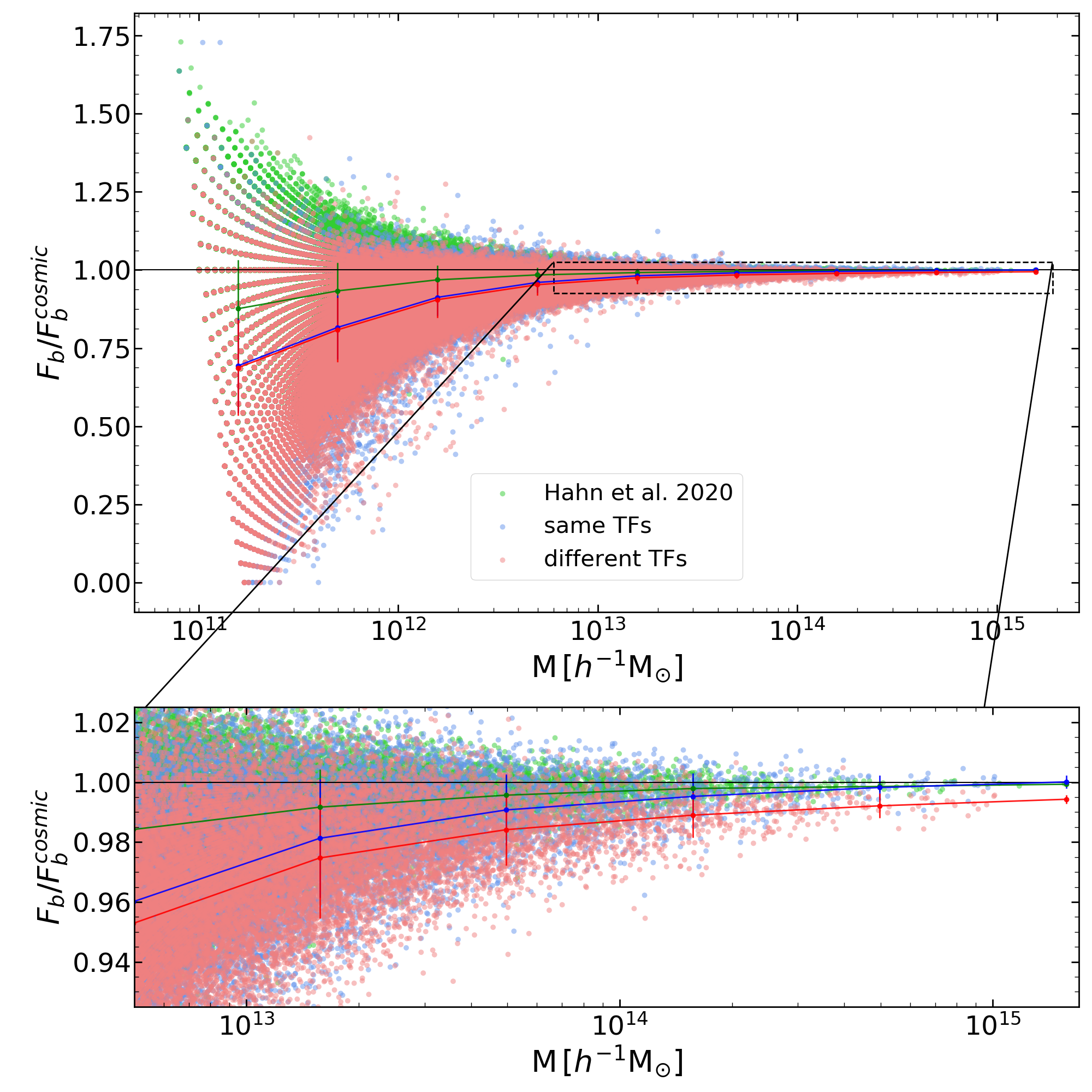}
\caption{\textbf{Upper panel:} baryon fraction of halos in our 2-fluid simulations at $z=0$ normalized to the cosmic mean, $F_b/F_b^{\rm cosmic}$. Each point represents an individual halo while the linked points with errorbars show the mean and $1\sigma$ error on the mean in mass bins. We show results for the 2-fluid-diff-TF and 2-fluid-same-TF simulations in red and blue respectively. We see that in both cases $F_b$ is consistent with the cosmic mean for well resolved halos of mass $M > 5 \times 10^{12} M_\odot / h$. We attribute the large downturn at small mass to AGS for baryons. To confirm that, in green we present results obtained from a simulation of \cite{Hahn:2020} who do not use AGS. We see that in this case the mean baryon fraction stays consistent withe cosmic mean at all mass. See text for more details. \textbf{Lower panel:} Same as upper panel but zoomed on the region of well resolved halos with mass $5 \times M > 10^{12} \, h^{-1} M_\odot$. We see that for these objects $F_b$ is indeed $95\%$ of the cosmic mean, and reaches the cosmic value for all three numerical setups at very high mass, although a small $\sim 1\%$ difference remains for the 2-fluid-diff-TF setup.}
\label{fig:fb-subplot}
\end{figure}

We first focus on the baryon fraction in halos $F_b$ (normalized to the cosmic mean $F_b^{\rm cosmic}$) as a function of total halo mass on \reffig{fb-subplot}. The red points in the upper panel show the baryon fraction of main halos identified in the first realization of our 2-fluid-diff-TF simulation (detailed in subsection \ref{sec:nbody}). We compute the baryon fraction as 
\be
F_b=\dfrac{M_{\rm baryon}}{M_{\rm total}},
\label{eq:FBb}
\ee
where $M_{\rm baryon}$ is the mass of baryons and $M_{\rm total}$ is the total mass of the halo (i.e. baryons + CDM). Notice that the mean cosmic baryon fraction for our cosmology is $F_b^{\rm cosmic}=0.1575$. We have a total sample of $282156$ halos at z=0, represented by individual points. We further compute the mean and $1\sigma$ error on the mean in each of our 8 mass bins indicated by the linked points with errorbars. Noticeably this figure shows that for high mass objects the baryon fraction is independent of mass, and that the mean baryon fraction within the virial radius is approximately equal to the cosmic mean, with relatively small scatter (the standard deviation is less than $3\%$ for halos of mass $M > 10 ^{13} \, h^{-1} M_{\odot} $), as can be seen on the lower panel of \reffig{fb-subplot}. 

The baryon fraction in halos and clusters was extensively studied in hydrodynamical simulations (e.g. \cite{He:2005hb,Ettori:2005hp,Kravtsov:2005ab,Crain:2006sb,Gottloeber:2007eb}). These works focused on late-time baryonic effects and we do not attempt to quantitatively compare our results with theirs here. However it is interesting to notice that all these works found a baryon fraction of roughly 0.9 of the cosmic mean in well resolved halos\footnote{Notice however that when feedback is included (which was not the case in these works) this fraction drops to roughly 0.6 of the cosmic mean, see e.g. figure 4 and 5 of \cite{Villaescusa-Navarro:2020rxg}.}. While we also find a baryon fraction slightly smaller than the universal one, in our case $F_b/F_b^{\rm cosmic}$ is greater than 0.95 for all halos with $M > 5 \times 10^{12} M_\odot / h$, and tends to unity at higher mass. This indicates that late-time baryonic effects are dominating over early ones for this quantity. However, depending on the precise considered mass, ignoring the effect of early baryon-CDM perturbations might lead to a non-negligible bias (up to $\sim 5\%$) when using the baryonic content of cluster to, e.g., infer the cosmic baryon fraction.

A noticeable decrease in $F_b$ is observed below $M \approx 5 \times 10 ^{12} \, h^{-1} M_{\odot} $. However this may be ascribed to our use of AGS for baryons and not actually be physical. Indeed the number of particles in these halos is of the order of the number of neighbors used to establish the smoothing length for baryons, and hence these halos fall in the regime where the computation of the trajectories of the baryon particles is not accurate. This may lead to an underestimation of the density of baryons in these halos. Notice that such a downturn was already observed in \cite{Crain:2006sb} who also attributed it to the poor resolution of low mass halos. To further test this we ran a hybrid test of 2-fluid simulations with the same initial transfer functions for different species (2fluid-same-TF simulations), showed in blue on \reffig{fb-subplot}. Again points show our individual halos in our total sample of $281309$ halos in this simulation, while the linked points with errorbars show the mean and error in each mass bin. We see that the mean value is not affected and that the scatter around the mean is also very similar in both cases which comforts our idea that the departure from the cosmic mean is not physical but merely a numerical artifact due to AGS, and gives us an indication of the limitations of our numerical setup. Finally, in order to confirm our hypothesis we use a simulation from \cite{Hahn:2020} (Hahn et al. (2020) in the following) who vary individual particle masses to circumvent the use of AGS. Their cosmology is very similar to ours\footnote{Only $\Ob=0.04897$ and $\sigma_8=0.8102$ differ from our parameters}, and we use a simulation with the same box size, number of particles, and initialized with the Zel'dovich approximation too. The results for $369933$ distinct halos as well as the mean relation are shown in green on the upper panel of \reffig{fb-subplot}. We can see that the deviations from the universal baryon fraction is much less than in our simulations, while the scatter stays approximately the same, confirming that the large downturn at low mass is due to AGS for baryons. 

\begin{figure}
\centering
\includegraphics[scale=0.25]{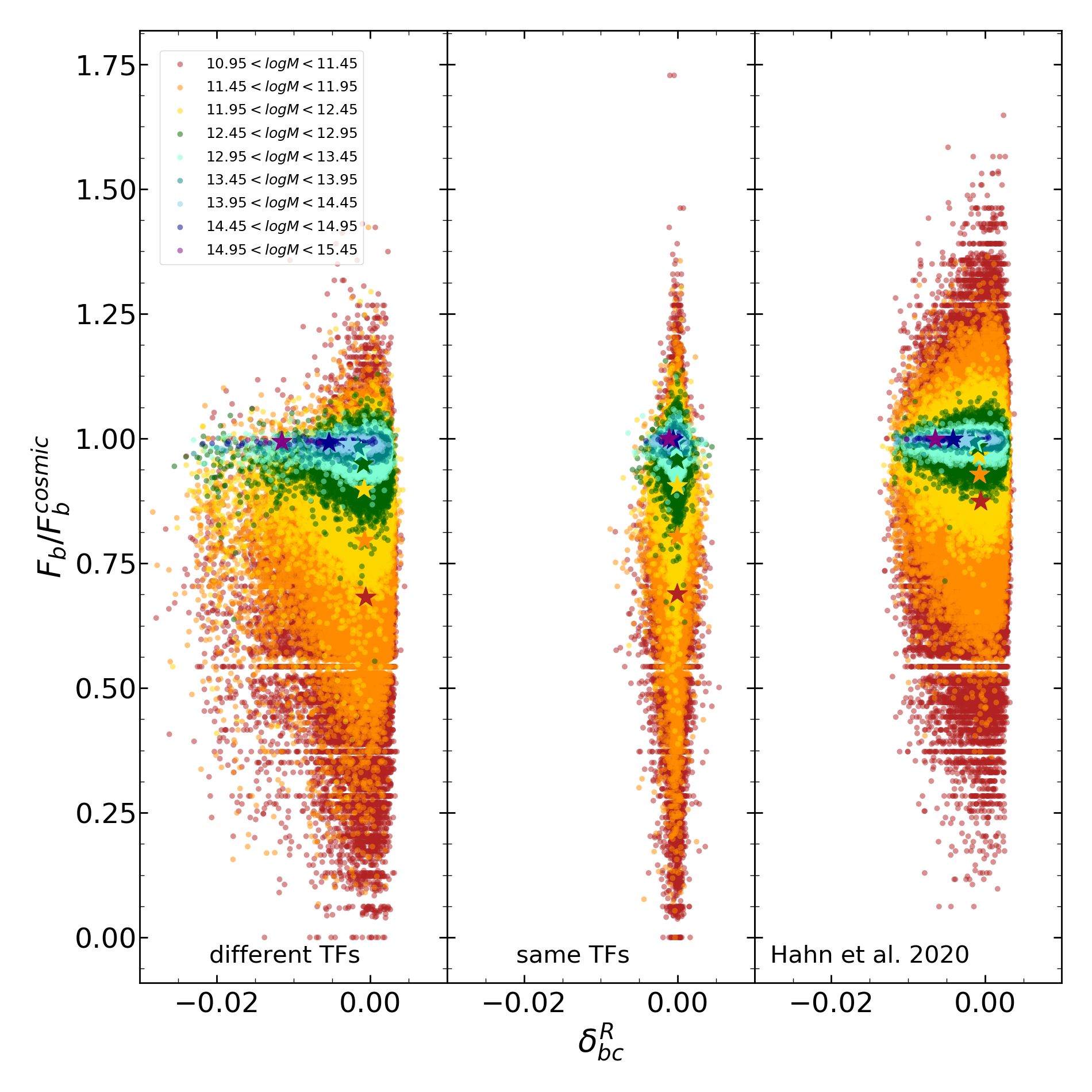}
\caption{The baryon fraction as a function of the baryon-CDM density smoothed on a scale $R=20 \Mpch$, $\delta^R_{bc}$, at $z=0$. The color code indicates different mass bins. The left, middle and right panels show results for the 2-fluid-diff-TF, 2-fluid-same-TF, and Hahn et al. 2020 simulations respectively. Each point represents a halo while star markers show the mean value.  We see that the scatter in $\dbc^R$ is relatively independent of mass and that the one in $F_b$ is independent of $\dbc^R$. The values of $\dbc^R$ are much smaller in the case of same transfer functions, and the mean values are close to zero, as expected. The fact that $\dbc^R$ is not exactly zero in that case results from numerical imprecision during the N-body evolution. The scatter in Hahn et al. (2020) seems roughly twice smaller than for our 2-fluid-diff-TF simulations. We however checked that this is only due to a few outliers with a high $\dbc^R$, while the mean values are similar between the rightmost and leftmost panels.}
\label{fig:dbcfb}
\end{figure}

Although on average the baryon fraction in halos is not affected by baryon-CDM perturbations, deviations with respect to the mean could be related to such fluctuations. Thus, we next turn to investigating the observed scatter in $F_b$ and its correlation with the local baryon-CDM perturbation $\dbc$. To do this we smooth $\dbc$ with a tophat filter on a scale of $20 \Mpch$ and interpolate it at the halo positions. This yields the environment baryon-CDM density of halos $\dbc^R$ that we can then plot against $F_b$. Results are presented on \reffig{dbcfb} for individual halos, color coded by mass. Star markers show the mean in each mass bin. The left panel represents the $F_{b} - \delta^R_{bc}$ plane for our 2-fluid-diff-TF simulation, the middle one displays results for the case of our 2fluid-same-TF simulation, and the right panel comes from the 2-fluid simulation communicated by Hahn et al. 2020 \cite{Hahn:2020} who use varying individual particle masses instead of AGS. 

We see that the scatter in $\dbc^R$ is relatively independent of the halo mass in all cases. The values taken by $\dbc^R$ are in majority negative, as expected from \reffig{density}, and they are much smaller in the case of the same transfer functions simulation, which is also expected. The fact that $\dbc^R$ is not exactly zero in that case results from numerical imprecision during the N-body evolution, and shows us what fraction of $\dbc$ is actually due to primordial baryonic effects compared to simple time evolution. We further notice that the scatter in $F_b$ seems independent of the one in $\dbc^R$. This indicates a small correlation between the local large-scale density $\dbc$ and the resulting baryon fraction in halos, confirming that the large scatter in $F_b$ at low mass is rather unphysical (i.e. $\dbc^R$ only weakly affects the measured $F_b$) but is due to numerical effects of poorly resolved halos. The scatter in Hahn et al. (2020) seems roughly twice smaller than for our 2-fluid-diff-TF simulations. We however checked that this is only due to a few outliers with a high $\dbc^R$, while the mean values are similar between the rightmost and leftmost panels. 


\subsection{2-fluids power spectra and halo bias}
\label{sec:2fPSbias}

We now turn to results for the impact of isocurvature baryon-CDM perturbations on halo clustering. We first focus on the correlations between $\d_m$, $\d_h$, and $\d_{bc}$ before presenting measurements of the associated bias $\bdbc$ obtained as explained in \refsec{measurebdbc} (equation \ref{eq:bdbc1f}). 

\begin{figure}
\centering
\includegraphics[scale=0.3]{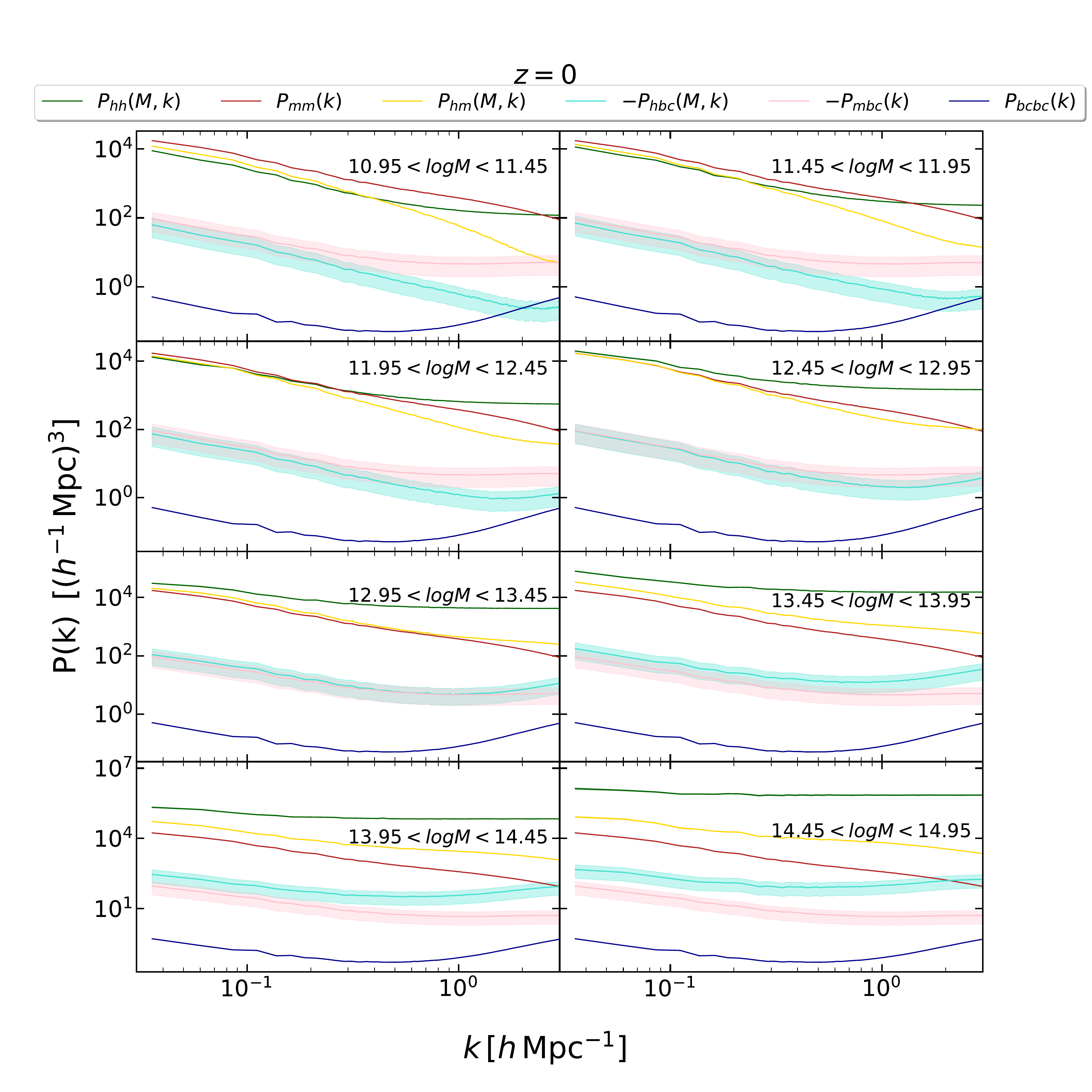}
\caption{2-fluid auto- and cross-power spectra constructed from $\d_m$, $\d_{bc}$ and $\d_h$ for all halo mass bins at redshift $z=0$. We plot $-P_{\rm mbc}$ and $-P_{\rm hbc}$ since these quantities are negative, reflecting the anticorrelation between $\d_{bc}$ and $\d_h$. The curves show the mean value and the shaded area show the $1\sigma$ error over all realizations of the 2-fluid-diff-TF simulations. We plot $P_{mm}$, $P_{mbc}$ and $P_{bcbc}$ on all panels even though they do not depend on mass to allow for a better visualization of the evolution of the halo power spectra with mass. The fact that $P_{hbc}$ is nonzero at all masses demonstrates that baryon-CDM perturbations $\d_{bc}$ affect the clustering of halos even in the low-redshift Universe. See text for more details.}
\label{fig:2fPS}
\end{figure}

\refFig{2fPS} shows the cross- and auto-power spectra between the halo field $\delta_h$, the matter field $\d_m$, and the baryon-CDM perturbation one $\dbc\approx \d_b - \d_c$ measured from our 2-fluid-diff-TF simulations at $z=0$. Each panel presents a different mass bin. $P_{hh}$, $P_{mm}$ and $P_{bc\,bc}$ are the halo, total matter, and relative baryon-CDM density perturbation auto-power spectra respectively, while $P_{hm}$, $P_{mbc}$, and $P_{hbc}$ are the halo-matter, matter-$\d_{bc}$, and halo-$\d_{bc}$ cross-spectra, respectively. We show $P_{mm}$, $P_{mbc}$ and $P_{bcbc}$ on all panels even though they do not depend on mass to allow for a better comparison of the evolution of the halo power spectra with mass. The lines show the mean of all realizations while light shaded areas show the $1\sigma$ error on the mean. We note that since $P_{hbc}$ and $P_{mbc}$ are negative, we plotted $-P_{hbc}$ and $-P_{mbc}$. These negative values reflect the anticorrelation between $\d_{bc}$ and $\d_h$, or $\d_{bc}$ and $\d_m$ coming from the fact that photon pressure in the early Universe prevents the collapse of baryons at early times, which in turn acts against the collapse of dark matter and the formation of halos. The fact that $P_{hbc}$ is nonzero at all masses demonstrates that baryon-CDM perturbations $\d_{bc}$ affect the clustering of halos even in the low-redshift Universe, and might have to be taken into account in the studies of structure formation. To our knowledge this is the first time that such correlations are shown. 

Turning to the mass evolution of the spectra involving $\d_h$, $P_{hh}$, $P_{hm}$, and $P_{hbc}$, we note that all increase with increasing halo mass. In particular $P_{hm}$ and $P_{hh}$ have the well-known behaviour of going from values smaller than $P_{mm}$ at low mass to dominating all the other spectra at high mass, reflecting the fact that low mass halos have a linear bias $b_1 < 1$ while more massive, rarer objects are highly biased (e.g. \cite{Lazeyras:2015} and references therein). We also see that at high mass the halo auto-power spectrum becomes roughly constant over the whole $k$ range indicating that it is dominated by the shotnoise term which is more important for these bigger and sparser objects. Since we have seen that the baryon-CDM field $\d_{bc}$ is smaller than the total matter one we expect $P_{bcbc}$, $P_{hbc}$ and $P_{mbc}$ to be smaller than $P_{hh}$, $P_{hm}$ and $P_{mm}$. In particular we expect  $P_{bcbc}$ to have the smallest value in comparison to the others. It is indeed what we observe on \reffig{2fPS}, and it is interesting to see that $P_{bcbc}$ is nonzero, while $P_{hbc}$ follows a similar mass evolution with respect to $P_{mbc}$ than $P_{hm}$ with respect to $P_{mm}$.

\begin{figure}
\centering
\includegraphics[scale=0.43]{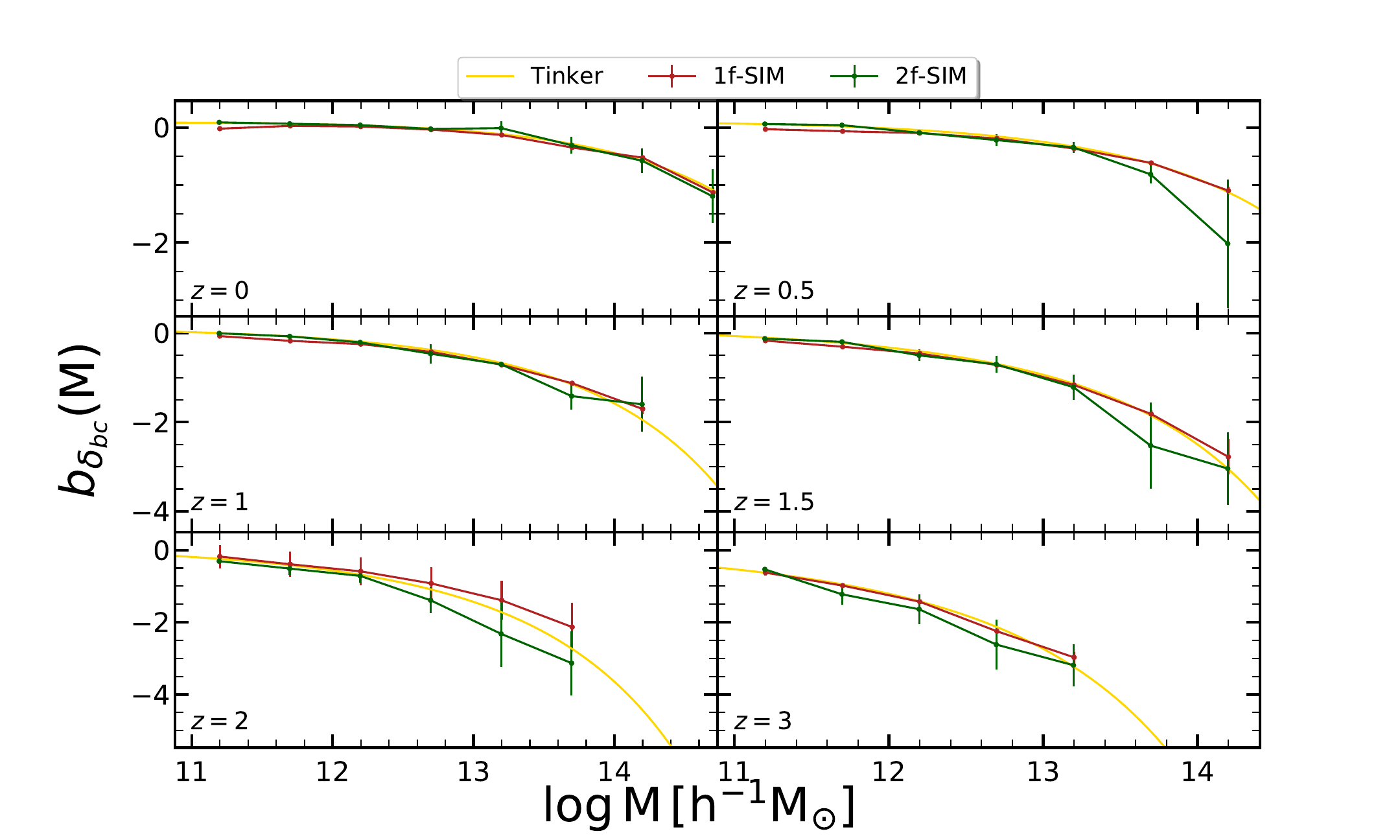}
\caption{Isocurvature baryon-CDM perturbations bias $\bdbc$ as a function of total halo mass for different redshift. The green dots joined by the solid line indicate $b_{\delta_{bc}}$ as measured in 2-fluid simulations using equation \ref{eq:bdbc1f}. The red dots joined by the solid line show the same measurement from 1-fluid separate universe simulations using equation \ref{eq:bias-separateuniv}. The errorbars show the $1\sigma$ error on the mean. The yellow solid line represents the Tinker prediction obtained using \refeq{bdbctinker}. We see that we get good agreement between the two methods as well as with Tinker. We find the same behaviour for this parameter as in previous work, i.e. it is overall negative with decreasing amplitude as a function of redshift, and more massive halos are more biased than small ones. See text for more details.}
\label{fig:bdbc-allz}
\end{figure}

From the results presented on \reffig{2fPS} we can obtain the bias associated to isocurvature baryon-CDM perturbations $\bdbc$ using equation \ref{eq:bdbc1f}. In order to maximize the signal to noise we follow the following procedure. We first obtain $b_1$ from our fiducial 1-fluid simulation by fitting a second order polynomial to the ratio in \refeq{b11f}, i.e. 
\be
\frac{P_{hm}^{\rm 1f}(k)}{P_{mm}^{\rm 1f}(k)} = b_1^{\rm 1f} + A_1 k^2,
\label{eq:phmopmm1f}
\ee
up to a maximum wavenumber $k_{\rm max}$. Here $A_1$ is simply the amplitude of the $k^2$ term we add to push the fit to higher $k_{\rm max}$. We do this for each of the first 4 realizations and for each mass bin. We then insert the result in the ratio \refeq{bdbc1f} that we also fit with a second order polynomial, i.e.
\be
\frac{P_{hm}(k)-b_1^{\rm 1f}P_{mm}(k)}{P_{mbc}(k)} = \bdbc + A_{bc} k^2,
\label{eq:phmmpmmopmbc}
\ee
where again $A_{bc}$ is an amplitude that we do not try to constraint. Importantly we use the result of $b_1^{\rm 1f}$ in realization $i$ to obtain $\bdbc$ in realization $i$ before averaging over all realizations and obtaining errorbars, and we do this for each mass bin. This allows us to cancel some of the cosmic variance since the CDM particles positions in our 1- and 2-fluid simulations are initialized with the same random seed. Since our simulations are of modest size (the fundamental mode $k_F$ is $0.025 \iMpch$) we choose a $\kmax=0.21 \iMpch$. We have tested the stability of our results under a change of this value and found it to be the optimal choice to maximize our signal to noise ratio while still obtaining unbiased results. Finally, in order to optimize the fit we put a loose mass-dependent flat prior on the value of $\bdbc$ of roughly ten times its amplitude at a given mass.

The results for $\bdbc$ as a function of halo mass are presented on \reffig{bdbc-allz} in green for various redshifts between 0 and 3. We further compare them to the results obtained from the 1-fluid ``separate universe simulations'' in the manner of \cite{Barreira:2019} in red. Since these authors described their procedure in great details, we only do a brief recap of it in \refapp{baryonCDMSU}. For each set of points the errorbars show the $1\sigma$ error on the mean obtained from all realizations. The yellow solid line represents the prediction obtained using the universal Tinker mass function \cite{Tinker:2008ff} as 
\be
b_{\delta}^{bc,univ} (z,M) = \dfrac{1}{\delta_{bc}}\Bigg[ \dfrac{n^{\rm SepUni,univ} (z,M)}{n^{\rm Fiducial,univ} (z,M) }- 1 \Bigg],
\label{eq:bdbctinker}
\ee
where $n^{\rm SepUni,univ}$ and $n^{\rm Fiducial,univ}$ are the universal halo mass function predictions computed with the Tinker fitting function and the linear matter power spectrum of the fiducial and separate universe cosmologies, respectively, and $\delta_{bc}$ is obtained with \refeq{dbcsepuni}. Notice that \refeq{bdbctinker} simply corresponds to \refeq{bias-low-high} applied to the Tinker mass function. 

Our results from the 2-fluid-diff-TF simulations are in overall very good agreement with the 1-fluid ones. It is an important cross-check since the two methods are completely independent and this parameter was only measured once before in \cite{Barreira:2019}. We observe the same behaviour as these authors, i.e. that $b_{\delta_{bc}}$ is negative over most of the mass and redshift range and is a decreasing function of halo mass. It can be seen that $b_{\delta_{bc}}$ decreases more with halo mass at higher redshift. $b_{\delta_{bc}}$ is negative at $z>1$, the only positive values for $b_{\delta_{bc}}$ being at $z=0$ for halo masses between $10^{10.95} < M < 10^{12.45}\,\, [ h^{-1} M_\odot]$, and at $z=0.5$ for halo masses between $10^{10.95} < M < 10^{11.95}\,\, [ h^{-1} M_\odot]$. The fact that $\bdbc$ is negative again reflects the fact the baryon-CDM perturbations make halo formation more difficult. Finally, our results are also in agreement with the curves derived from the Tinker mass function.

We also note that even though we neglected the velocity bias $b_{\theta_{bc}}\theta_{bc}$ in \refsec{measurebdbc}, our measurements of $\bdbc$ from 2-fluid simulations are in complete agreement within the errorbars with those from 1-fluid separate universe simulations. This means that indeed $\theta_{bc}$ is subdominant at low redshift, as could be expected since it is a decaying term, and as was already pointed out in e.g. \cite{Schmidt:2016, Barreira:2019}. This also mean that in order to try and measure this term from 2-fluid simulations one would need a much larger volume in order to obtain a detection. This goes beyond the scope of this paper and we defer it to future work.   

\begin{figure}
\centering
\includegraphics[scale=0.35]{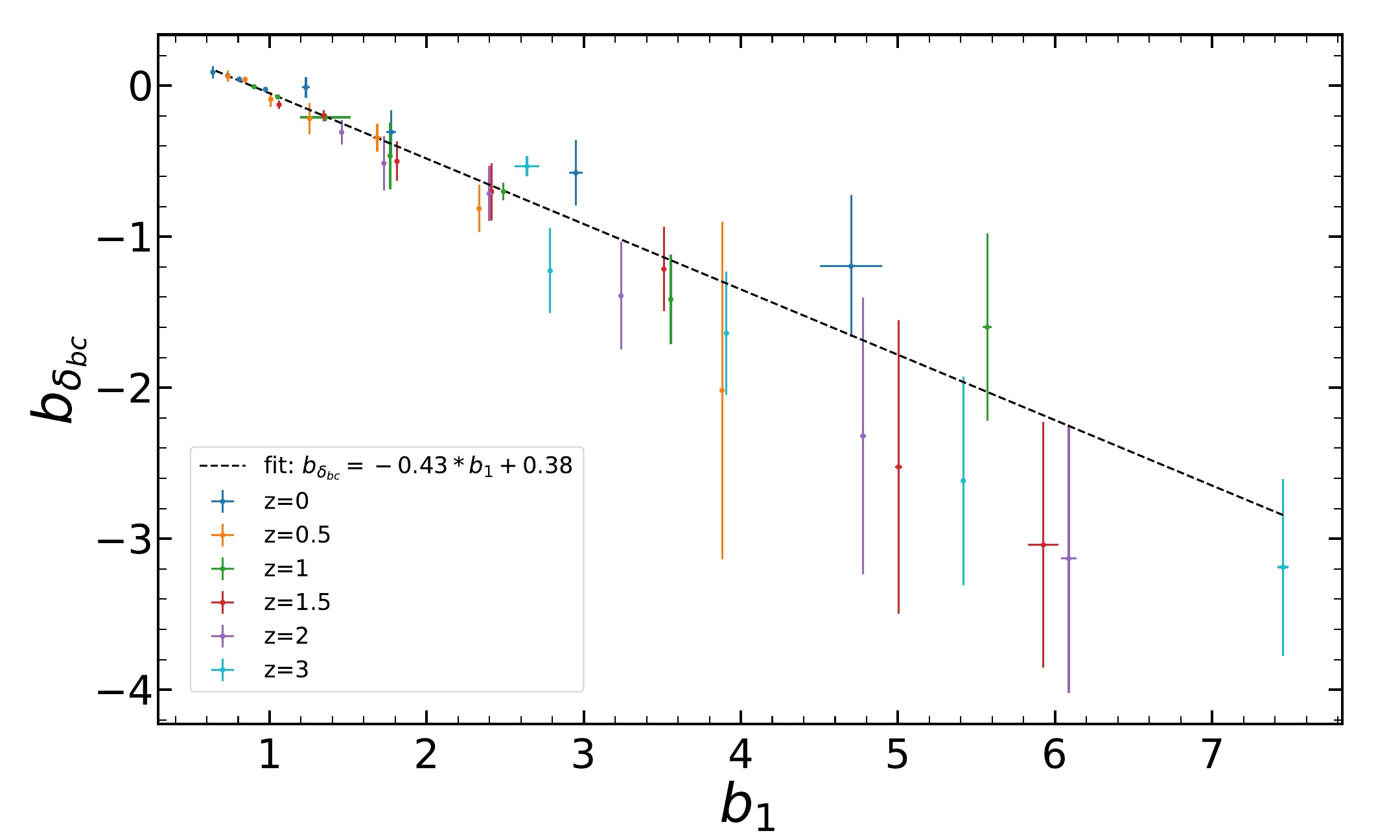}
\caption{Isocurvature baryon-CDM perturbations bias $\bdbc$ as a function of the linear bias $b_1$ at different redshift indicated by the color coding. We see no clear trend with redshift although results at $z=0$ seems to be systematically higher than the other ones. The dashed line is a linear fit obtained by fitting all points simultaneously.}
\label{fig:bdbc-b1}
\end{figure}

Finally on \reffig{bdbc-b1} we present $b_{\delta_{bc}}$ as a function of the linear bias $b_{1}$ at different redshift indicated by the color coding. We see no significant trend with redshift although results at $z=0$ seems to be systematically higher than the other ones, especially at high mass as we noticed on \reffig{bdbc-allz}. This figure shows us an approximately linear behaviour between these two quantities but while $b_{1}(z,M)$ is always positive $b_{\delta_{bc}}(z,M)$ is majoritarly negative. This motivated us to provide a fitting formula for this relation which might prove useful for accurate modeling of the halo power spectrum without introducing a new free parameter. We choose a linear fit, shown as the dashed line on \reffig{bdbc-b1}, given by
\be
\bdbc(b_1) = -0.43 b_{1} + 0.38.
\label{eq:fitbdbc}
\ee

From the results on \reffigs{2fPS}{bdbc-b1} we can assess the importance of the contribution of $\dbc$ to the halo power spectrum by comparing its value when including it or not, i.e
\be
P_{hh} = b_1^2 P_{mm} \; \; \; {\rm vs} \; \; \; P_{hh} = b_1^2 P_{mm} + 2 b_1 \bdbc P_{mbc} + \bdbc^2 P_{bcbc}.
\label{eq:phhcomp}
\ee 
We do this comparison at $k=0.1 \iMpch$ and at $z=0$ for all mass bins and find a maximum effect of roughly $0.3\%$ for the highest mass halos\footnote{Notice that for a pure affine relation between $\bdbc$ and $b_1$ one could reabsorb the effect of $\dbc$ on the halo power spectrum in a scale dependent linear bias (by neglecting the last term in \refeq{phhcomp}) as is done for neutrinos. The fact that $\bdbc$ is nonzero at $b_1=0$ prevents from doing so.}. Notice that at low mass the contribution due to $\dbc$ is negative while it becomes positive at high masses. For a more concrete case we consider a Euclid-like survey at redshift $z=1$ for which the linear bias should be of 1.46. We can then use \refeq{fitbdbc} to infer the value of $\bdbc= -0.248$. Plugging everything into \refeq{phhcomp} we get a relative difference of $0.3\%$. Doing the same exercise for a higher redshift sample such as the DESI QSO one (centered around $z=3$ with $b_1=3$ and hence $\bdbc=-0.91$) yields a $1\%$ impact of baryon-CDM perturbations on $P_{hh}$ at the same scale. 

\begin{figure}
\centering
\includegraphics[scale=0.4]{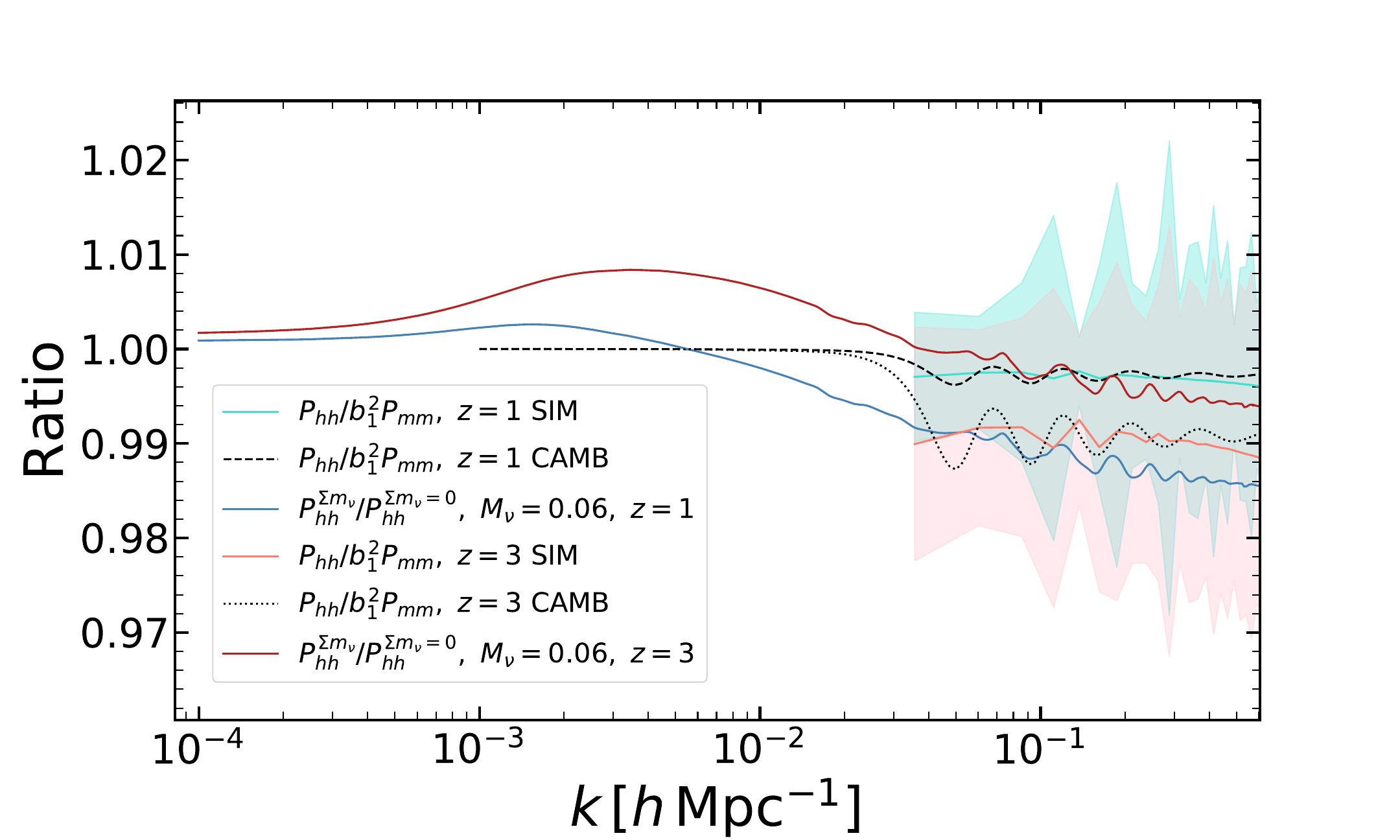}
\caption{Comparison between the effect of early baryon-CDM perturbations and the effect of massive neutrinos on the halo power spectrum $P_{hh}$ for a Euclid-like survey at $z=1$, and Quasars from a DESI-like survey at $z=3$. The light blue and light red curves with shaded regions for errorbars represent the ratio of the two models in \refeq{phhcomp} from our simulations at $z=1$ and $z=3$ respectively. The dashed and dotted black curves show the same quantity when we use power spectra from CAMB at $z=1$ and $z=3$ respectively in \refeq{phhcomp}. The blue and red curve show the effect of massive neutrinos of total mass $\Sigma M_{\nu}=0.06 \, eV$ for a Euclid-like and DESI QSO-like survey at $z=1$ and $z=3$ respectively. We see that while the effect of neutrinos clearly dominates at lower redshift, it is subdominant compared to that of $\dbc$ at $z=3$.}
\label{fig:phhcomparison}
\end{figure}

We put these values in contrast with the effect expected from massive neutrinos for these two samples. We consider a total neutrino mass of $\Sigma M_\nu=0.06 eV$ consistent with the lower limit set by neutrino oscillations experiments (e.g. \cite{deSalas:2018bym} and references therein). We consider the scale-dependent effect induced by neutrinos both on the linear matter power spectrum $P_L$ and the linear bias $b(k)$ as parametrized in \cite{Chiang:2018laa}, i.e 
\be
\frac{P_{hh}^{\Sigma M_\nu}(z, k)}{P_{hh}^{\Sigma M_\nu=0}(z,k)} = \frac{[1+(b_1(z)-1)f(k)]^2P_L^{\Sigma M_\nu}(z,k)}{b_1(z)^2P_L^{\Sigma M_\nu=0}(z,k)},
\label{eq:phhnu}
\ee
where the superscripts ``$\Sigma M_\nu$'' and ``$\Sigma M_\nu=0$'' stand for the model with and without neutrinos respectively, and $f(k)$ represents the transition between scales larger and smaller than $k_{\rm fs}$ (the neutrino free-streaming scale). It can be approximated by a step-like $\tanh$ function with a width of $\Delta_q$ \cite{Munoz:2018ajr}
\be
f(k) = \left[1+\frac{3\Delta_L}{2}\times \left(\tanh\left(\frac{q}{\Delta_q}\right)+1 \right)\right]
\label{eq:fk}
\ee
where $q = \log(5k/k_{\rm fs})$, $\Delta_L = 0.55f_\nu$ with $f_\nu=0.045$ the neutrino fraction for our choice of $\Omega_m$ and $\Sigma M_\nu$, and we used $\Delta_q = 0.6$. Finally we compute the ratio of the linear power spectra from CAMB by keeping the total matter density $\Omega_m$, the baryon fraction $f_b$ and the initial amplitude of perturbation $A_s$ fixed. 

Result are presented on \reffig{phhcomparison} for a Euclid-like galaxy sample and a DESI QSO one. The light blue and light red curves with shaded regions for errorbars represent the ratio of the two models in \refeq{phhcomp} from our simulations at $z=1$ and $z=3$ respectively, while the dashed and dotted black curves show the same quantity when we use power spectra from CAMB. The blue and red curve show the effect of massive neutrinos of total mass $\Sigma M_{\nu}=0.06 \, eV$ at those two redshifts. We see that while the effect of neutrinos clearly dominates at lower redshift, it is subdominant at $z=3$. Hence the impact of $\dbc$ has to be taken into account if one is to put solid constraint on neutrino mass from DESI QSO-like sample, which is a fundamental point for future surveys.


\section{Summary and conclusions}
\label{sec:conclusion}

In this paper we have performed 2-fluid cosmological N-body simulations to study the impact of baryon-CDM isocurvature perturbations on dark matter halos. Particularly we concentrated on the baryon fraction in halos $F_b$ as a function of halo mass and local baryon-CDM relative density, and the cross-correlation between the halo field and the baryon-CDM perturbation field. We also measured the associated baryon-CDM bias parameter $b_{\delta_{bc}}$, and compared our results from 2-fluid simulations with ones obtained from 1-fluid separate universe technique as performed in \cite{Barreira:2019}. We also assessed the impact of $\dbc$ on the halo power spectrum for a Euclid-like galaxy sample and a DESI Quasar one and compared it to that of massive neutrinos. Since the setup for these 2-fluid simulations is nontrivial we made some numerical tests to show the validity of our simulations in section \ref{sec:numtests}. A critical point is the need to use AGS for the baryon fluid in order to retrieve agreement with linear theory on large scales, although other numerical techniques allow to avoid it \cite{Bird:2020,Hahn:2020}.

Our main findings can be summarized as follows:
\begin{itemize}
     \item The baryon fraction in halos is slightly smaller than the universal one and is a weakly dependent function of halo mass. For mass $5 \times M > 10^{12} h^{-1} (M_\odot)$ we found it to be larger than $95\%$ the cosmic mean with relatively small scatter. We measured a noticeable downturn in lower mass bins for both our 2-fluid-diff-TF and 2-fluid-same-TF simulations, which we ascribe to our softened forces. 
     \item To further study the scatter in $F_b$ we looked at its correlation with the local large-scale baryon-CDM relative density, and found it to be small . This confirmed that the large scatter in $F_b$ at low halo mass is rather unphysical (i.e. $\dbc^R$ only weakly affects the measured $F_b$), and is rather due to numerical effects of poorly resolved halos. 
     \item The halo-baryon-CDM cross-spectra $P_{h\,bc}$ is nonzero, showing that baryon-CDM perturbations affect the clustering of structures even at low redshift. $P_{hbc}$ and $P_{mbc}$ are negative reflecting the anticorrelation between $\d_{bc}$ and $\d_{h}$, or $\d_{bc}$ and $\d_{m}$. $P_{bcbc}$ has the smallest value in comparison to the other power spectra but it is nonzero. To our knowledge this is the first time such correlations are measured.
     \item Our results for $\bdbc$ from the 2-fluid simulations are in agreement with those from 1-fluid separate universe simulations. It is an important cross-check since the two methods are completely independent and this parameter was only measured once before in \cite{Barreira:2019}. This parameter is negative reflecting again the anticorrelation between $\dbc$ and halo formation. 
     \item Even though we neglected the effect of relative velocities to measure $\bdbc$ from 2-fluid simulations we found perfect agreement with previous results within the errorbars. This confirms that the term proportional to $\theta_{bc}$ is subdominant in the bias expansion.
     \item We found a linear relation between $\bdbc$ and $b_{1}$ at all $z$ and provided a fit in \refeq{fitbdbc}. 
     \item We found the contribution of terms proportional to $\bdbc$ to the halo power spectrum to be at maximum $0.3\%$ at $k=0.1 \iMpch$ at $z=0$. We compared this to the effect of massive neutrinos for a Euclid-like galaxy sample at $z=1$ and a DESI-like quasar one at $z=3$, finding the impact of $\dbc$ to be dominant in the latter case which is a fundamental point for future surveys.
\end{itemize}

2-fluid numerical simulations have proven to be an effective tool to study the impact of baryon-CDM relative perturbations on LSS. Several works have now worked out near optimal setups for these simulation \cite{Bird:2020, Hahn:2020} and have opened the door to study how various LSS observables are affected. While we showed that halo clustering is only weakly affected and that measuring the relative velocity bias $b_{\theta_{bc}}$ would require a very large simulation volume, we also showed that the baryon fraction in halo deviates from the cosmic mean by a non-negligible amount even with respect to late-time hydrodynamical effects for some halo masses. Finally we also showed that the effect of $\dbc$ can be degenerate with that of neutrinos for some future surveys. In the future it would be interesting to study how cosmic voids are affected by baryon-CDM perturbations, as well as the modulation of the BAO feature in such simulations since this is known to be a probe of primordial isocurvature perturbations \cite{Heinrich:2019sxl}.  

\acknowledgments{We thank Alex Barreira for useful discussions at the early stages of this work. HK, TL and MV are supported by INFN INDARK grant. MV also acknowledges contribution from the agreement ASI-INAF n.2017-14-H.0.  REA acknowledges the support of the ERC-StG number 716151 (BACCO). OH acknowledges support from the European Research Council under ERC StG. number 679145 (COSMO-SIMS).}

\appendix

\section{Baryon-CDM bias from 1-fluid ``separate universe simulations''}
\label{app:baryonCDMSU}

Here we describe in summary the separate universe technique for the case of baryon-CDM perturbations in 1-fluid simulations. As already discussed in \cite{Barreira:2019}, the effects of baryon-CDM density perturbations on structure formation can be mimicked by a change in the baryon density $\Omega_{b}$ and in the CDM one $\Omega_{c}$, keeping the total matter density $\Om$ constant. These changes can be described by a parameter $\Delta_{b}$ as $\Tilde{\Omega}_{b}=\Omega_{b}[1+\Delta_{b}]$ and $\Tilde{\Omega}_{c}=\Omega_{c}[1- f_{b} \Delta_{b}]$, where the tilde indicates the baryon and CDM density in the modified cosmology, and $f_{b}$ is here the ratio of the baryon density over the CDM density in the fiducial cosmology $f_{b}=\Omega_b/\Omega_c$\footnote{Note that it is not the baryon fraction here but we follow the notation of \cite{Barreira:2019}}. By using the following relation and a Taylor expansion we get 
\be
1+\delta_{bc}=\dfrac{\Tilde{\Omega}_{b}/\Tilde{\Omega}_{c}}{\Omega_{b}/\Omega_{c}}=\dfrac{1+\Delta_{b}}{1-f_{b}\Delta_{b}}\approx 1+(1+f_{b})\Delta_{b}.
\label{eq:dbcsepuni}
\ee
We follow the same procedure as \cite{Barreira:2019} and consider three different cosmologies dubbed Fiducial, High and Low with the following parameters: $\Delta_{b}^{\rm  High}=0.05$, $\Delta_{b}^{\rm Low}=-0.05$ and $\Delta_{b}^{\rm Fiducial}=0$, corresponding to $\Tilde{\Omega}_{b}^{\rm High}=0.0515$, $\Tilde{\Omega}_{c}^{\rm High}=0.2596$, and  $\Tilde{\Omega}_{b}^{\rm Low}=0.0466$, $\Tilde{\Omega}_{c}^{\rm Low}=0.2645$ while we keep $\Omega_{m}$ and all other cosmological parameters fixed in all three different cosmologies. We measure the baryon-CDM density bias $\bdbc$ from 16 realizations of 1-fluid simulations as follows:
\begin{align}
b_{\delta_{bc}}(z,M) = \dfrac{b_{\delta_{bc}}^{\rm High}(z,M) + b_{\delta_{bc}}^{\rm Low}(z,M)}{2},  
\label{eq:bias-separateuniv}
\end{align}
where
\begin{align}
b_{\delta_{bc}}^{\rm High}(z,M) = \dfrac{1}{\delta_{bc}^{\rm High}} \Bigg[ \dfrac{N^{\rm High}(z,M)}{N^{\rm Fiducial}(z,M)}-1 \Bigg], \nonumber \\ 
b_{\delta_{bc}}^{\rm Low}(z,M) = \dfrac{1}{\delta_{bc}^{\rm Low}} \Bigg[ \dfrac{N^{\rm Low}(z,M)}{N^{\rm Fiducial}(z,M)}-1 \Bigg].
\label{eq:bias-low-high}
\end{align}
Here $\delta_{bc}=(1+f_{b})\Delta_{b}$, and $N(z,M)$ denotes the number of halos found in the corresponding cosmology at redshift $z$ in some mass bin $M$. We refer the interested reader to \cite{Barreira:2019} for more details about this procedure.


\section{Additional numerical tests}
\label{app:tests}

\begin{figure}
\centering
\includegraphics[scale=0.35]{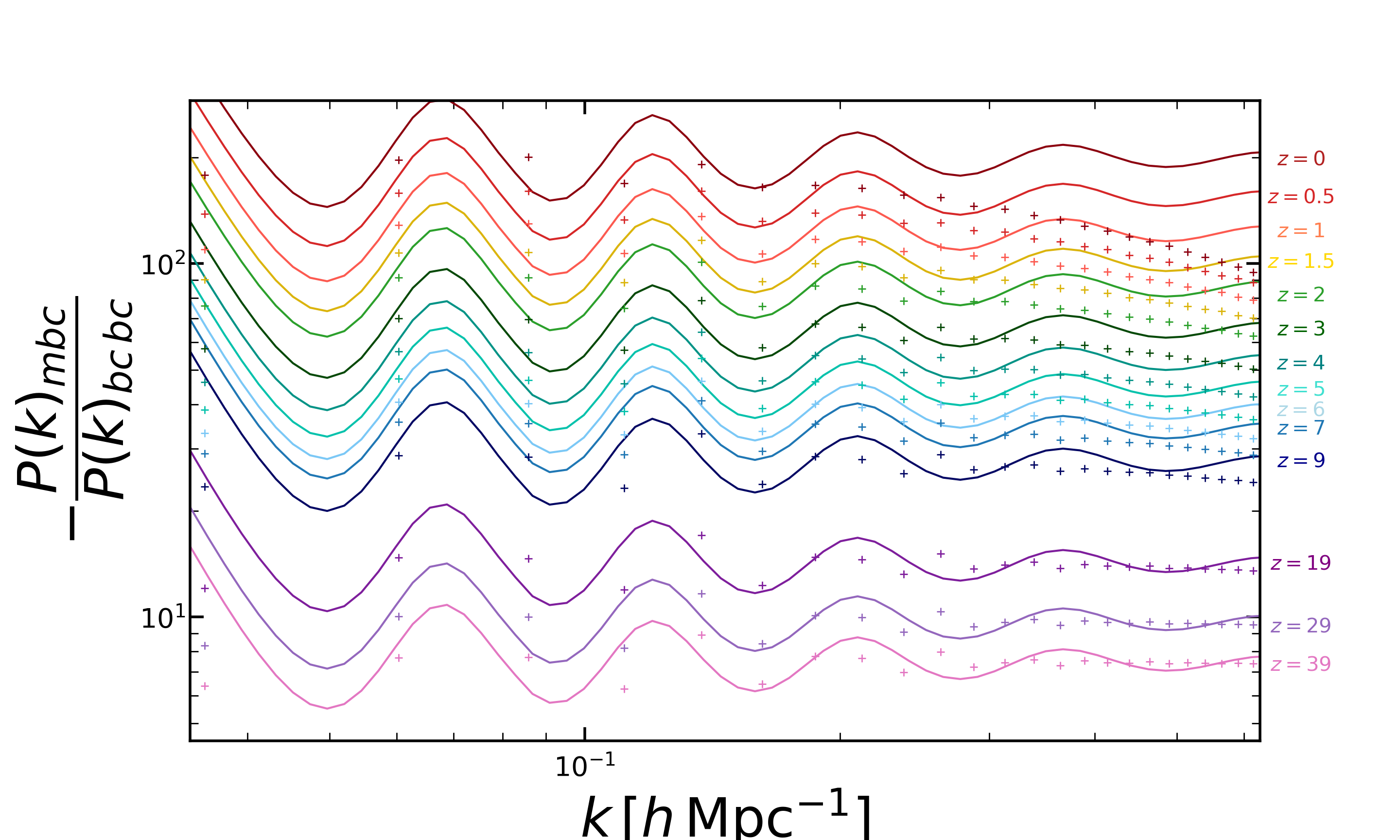}
\caption{Comparison between the measurement of the ratio $-P_{mbc}/P_{bcbc}$ from the 2-fluid-diff-TF simulations with the CAMB prediction. We present results for 14 different output redshifts indicated by the color coding. Solid lines show the prediction of linear perturbation theory and plus (``+'') markers represent simulations results. We find again overall good agreement validating once more our numerical setup. The suppression at small scales is due to our treatment of the force softening for baryons and nonlinear effects. The small inconsistency of our measurement  with the theoretical prediction on the largest scale is due to the small number of modes in this first $k$ bin.}
\label{fig:PmbcoverPbcbc}
\end{figure}

In this appendix we present additional numerical tests to validate our numerical setup. We first compare our measurement of the ratio $-P_{mbc}/P_{bcbc}$ with the linear prediction from CAMB at various redshift on \reffig{PmbcoverPbcbc}. The ``+'' markers shows the results from the 2-fluid-diff-TF simulations, and the solid lines show the expectation from linear perturbation theory. We find good agreement between the two at all redshift up to mildly nonlinear scales validating once more our setup. The small discrepancy observed on the largest scale is most likely due to the small number of modes in this lowest $k$ bin. As can be seen on \reffig{PmbcoverPbcbc}, the matter-$bc$  cross spectra, $P_{m\,bc}$ dominates the $P_{bc\,bc}$ auto power spectra in all output redshift.

\begin{figure}
\centering
\includegraphics[scale=0.4]{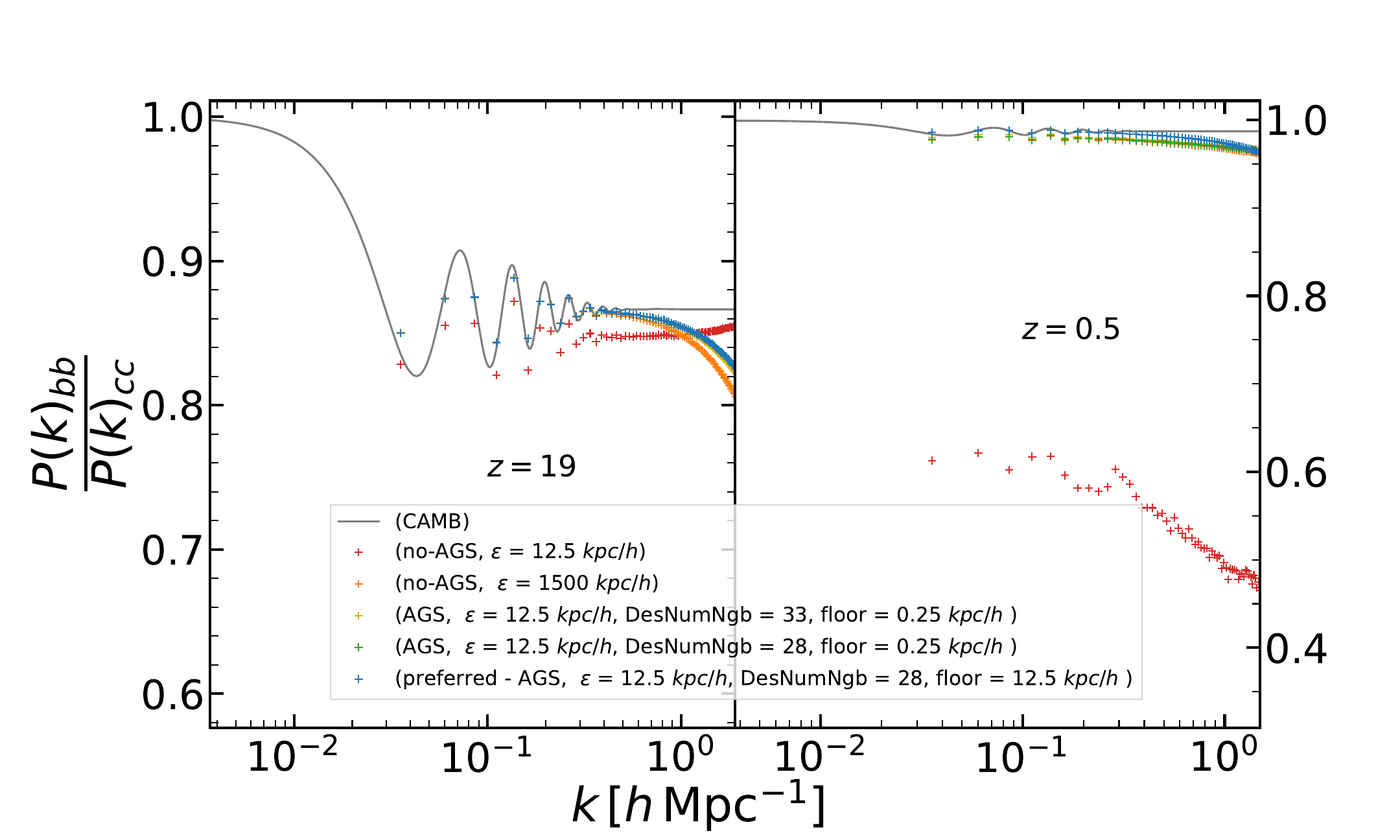}
\caption{Numerical tests of the effect of the force resolution in 2-fluid simulations. We compare the ratio of the power spectrum of baryons and CDM for five different runs with different baryon softening lengths at $z=19$ and $z=0.5$ on the left and right panel respectively. The gray line displays the CAMB expectation. Red plus markers show a test run using a Plummer-equivalent softening length, $\epsilon$, set to $1/40$-th of the mean inter-particle separation corresponding to $\epsilon = 12.5 {\rm kpc}/h$. Orange plus markers show a similar test but using a very large softening length for baryons, set to $3$ times the mean inter-particle separation corresponding to $\epsilon = 1500 {\rm kpc}/h$. These two test do not include AGS. Yellow plus markers show the results of another run using AGS, where forces between particles are softened adaptively using an SPH kernel with a width set by the distance to the $33^{\rm rd}$ closest neighbour. Green plus markers display a similar run but using the SPH kernel with a width set by the distance to the $28^{\rm th}$ neighbour. Both runs set a floor for the minimum softening length of baryon to $0.25 {\rm kpc}/h$. The blue markers show our preferred setup where the SPH kernel width is set by the distance to the $28^{\rm th}$ neighbour but the minimum allowed SPH smoothing length is raised to $12.5 \rm kpc /h$. All runs use a fixed softening length of $12.5 \rm kpc /h$ for CDM particles. We clearly see the need to use AGS with a reasonable softening length floor in order to recover the linear prediction from CAMB.}
\label{fig:tests}
\end{figure}

Finally, figure \ref{fig:tests} shows the relative difference in the clustering of baryons and CDM as the ratio $P_{bb}/P_{cc}$. This corresponds to the middle panels of \reffig{PbboverPc}. The left panel shows the ratio at $z=19$, and the right one at $z=0.5$. Here we plot results from different runs featuring different force resolutions, with and without AGS for baryons. Furthermore we show the impact of using different desired number of neighbours to setup the SPH kernel width (DesNumNgb), and different minimum allowed SPH smoothing length (floor) in case AGS is used. The desired number of SPH smoothing neighbours in the Gadget code represents the effective number of neighbours defined as the mass inside the kernel divided by the particle mass, and is kept constant very close to the desired value. A range can be defined to allow variation of the number of neighbours around the target value that we keep constant equal to 2. The floor is the minimum allowed SPH smoothing length which is used instead of the SPH kernel width in very dense regions. The color coding on \reffig{tests} indicates the particular run, and the CAMB expectation is shown by the solid gray line. As discussed in detail in \cite{Angulo:2013qp}, using adaptive gravitational softening for baryons seems to correctly recover the relative large-scale clustering of baryons and CDM while avoiding the use of a very large smoothing length for baryons. This fact can be seen in the left panel of figure \ref{fig:tests} at $z=19$ for instance. The test run denoted with red plus (``+'') markers shows results without using AGS but using a Plummer-equivalent softening length $\epsilon$ set to $1/40$-th of the mean inter-particle separation. Although this is a standard value used in state-of-the-art simulations, in this case, it is underestimating the strength of the coupling between our two particle species yielding results completely inconsistent (about $ \sim 5 \%$) with the linear theory expectation. As can be seen in the right panel, this inconsistency becomes more and more important as we move to lower redshift (here for instance at $z=0.5$ we can see $ \sim 40 \%$ discrepancy with the linear theory). To remove this discrepancy at least at high redshift we can use a very huge softening length, $\epsilon$, set to $3$ times the mean inter-particle separation only for the baryon particles. For the CDM particles, we still set it to $1/40$-th of the mean inter-particle separation. The result of this run is denoted in orange on \reffig{tests}. In spite of removing the discrepancy with the linear theory on large-scales, we can see a lack of power at smaller scales in this case. As can be seen on the right panel of figure \ref{fig:tests}, we still have about $\sim 3 \%$ discrepancy with the linear theory expectation on large-scales at $z=0.5$. The same occurs in another test, denoted in yellow, where we use AGS for baryons. In this test run, the forces between particles are softened adaptively using an SPH kernel with a width set by the distance to the $33^{\rm rd}$ neighbour, denoted as DesNumNgb=33. By decreasing the width of this kernel we can slowly reach the linear theory expectation even at lower redshift (green markers). Notice however that the minimum allowed SPH smoothing length (floor) was set to a very low values in these cases. By setting it to $1/40$-th of the mean inter-particle separation and setting DesNumNgb=28 we can remove the discrepancy at all the redshift up mildly nonlinear scales. This is our final setting denoted with blue markers on figure \ref{fig:tests}. It is in agreement with linear theory up to $k \sim 0.3 \, h \rm Mpc^{-1}$. we refer to this setup as 2-fluid-diff-TF and use it to obtain the results presented in the main text.


\FloatBarrier
\bibliographystyle{JHEP}
\bibliography{ref}

\end{document}